\documentclass[fleqn,usenatbib]{mnras}

\usepackage{newtxtext,newtxmath}

\usepackage[T1]{fontenc}
\usepackage{ae,aecompl}

\usepackage{graphicx}	
\usepackage{amsmath}	
\usepackage{amssymb}	
\usepackage{xspace}
\usepackage{gensymb}
\usepackage{hyperref}
\hypersetup{draft}

\defcitealias{Poggianti2017a}{Paper I}
\defcitealias{Bellhouse2017}{Paper II}
\defcitealias{Fritz2017}{Paper III}
\defcitealias{Gullieuszik2017}{Paper IV}
\defcitealias{Moretti2018}{Paper V}
\defcitealias{Vulcani2017c}{Paper VIII}
\defcitealias{Vulcani2018}{Paper VII}
\defcitealias{Jaffe2018}{Paper IX}

\newcommand{\Ha}{H$\alpha$\xspace}
\newcommand{\Hb}{H$\beta$\xspace}

\newcommand{\kms}{$\rm km \, s^{-1}$\xspace}
\newcommand{\kube}{{\sc kubeviz}\xspace}
\newcommand{\sinopsis}{{\sc sinopsis}\xspace}

\title[Physical processes in groups]{GASP. XII. 
The variety of physical processes occurring in a single galaxy group in formation} 

\author[B. Vulcani et al.]{Benedetta Vulcani,$^{1}$\thanks{E-mail: benedetta.vulcani@inaf.it (BV)}
Bianca M. Poggianti,$^{1}$
Yara L. Jaff\'e,$^{2}$
Alessia Moretti,$^{1}$
\newauthor
Jacopo Fritz,$^{3}$
Marco Gullieuszik,$^{1}$
Daniela Bettoni,$^{1}$
Giovanni Fasano,$^{1}$
\newauthor
Stephanie Tonnesen,$^{4}$
and Sean McGee$^{5}$
\\
$^{1}$INAF- Osservatorio astronomico di Padova, Vicolo Osservatorio 5, IT-35122 Padova, Italy\\
$^{2}$Instituto de F\'isica y Astronom\'ia, Universidad de Valpar\'iso, Avda. Gran Breta\~na 1111 Valpara\'iso, Chile \\
$^{3}$Instituto de Radioastronom\'ia y Astrof\'isica,
UNAM, Campus Morelia, A.P. 3-72, C.P. 58089, Mexico\\
$^{4}$Center for Computational Astrophysics, Flatiron Institute, 162 5th Ave, New York, NY 10010, USA\\
$^{5}$University of Birmingham School of Physics and Astronomy, Edgbaston, Birmingham, England}

\date{Accepted XXX. Received YYY; in original form ZZZ}

\pubyear{2018}

\begin{document}
\label{firstpage}
\pagerange{\pageref{firstpage}--\pageref{lastpage}}
\maketitle

\begin{abstract}
GAs Stripping Phenomena in galaxies with MUSE (GASP) is a program aimed at studying gas removal processes in nearby galaxies in different environments. We present the study of four galaxies that are part of the same group ($z=0.06359$) and highlight the multitude of mechanisms affecting the spatially resolved properties of the group members. One galaxy is  passive  and shows a regular stellar kinematics. The analysis of its star formation history indicates that the quenching process lasted for a few Gyr and that the star formation declined throughout the disk in a similar way, consistent with strangulation. Another galaxy is characterised by a two-component stellar disk with an extended gas disk that formed a few $10^8$ yr ago, most likely  as a consequence of gas accretion. The third member is a spiral galaxy at the edges of the group, but embedded in a filament. We hypothesise that the compression exerted by the sparse inter-galactic medium on the dense circum-galactic gas switches on star formation in a number of clouds surrounding the galaxy (``cosmic web enhancement''). Alternatively, also ram pressure stripping might be effective. Finally, the fourth galaxy is a spiral  with a truncated ionised gas disk and an undisturbed stellar kinematics. An analytical model of the galaxy's restoring pressure, and its location and velocity within the cluster, suggest ram pressure is the most likely physical mechanism in action. 
This is the first optical evidence for stripping in groups. 
\end{abstract}




\section{Introduction}
\label{sec:intro}

Observations show that in the local Universe the vast majority of the galaxies are found in groups \citep{Zabludoff1998}, which are associations of typically no more than a few dozens members with a total mass content of $\sim 10^{13}{\rm M_\odot}$.  Groups range from loose associations a few times denser than their surroundings to galaxy clusters, which are the largest virialised structures in the Universe, with a total mass of $\sim 10^{15}{\rm M_\odot}$, but containing only 1\% of all galaxies. 

In the current picture, dark matter structure formation proceeds in a hierarchical bottom-up sequence: small virialised haloes form first, and then grow by accretion and merging \citep{Huchra1982, Eke2004, Berlind2006, Knobel2009}. Groups therefore are considered important laboratories, as the study of their members can provide new understanding of the evolution of galaxies, of the large-scale structure, and of the underlying cosmological model.  

Galaxy evolution depends primarily on two key parameters: the environment,  and the stellar mass.
In the literature, huge effort has been dedicated to disentangling the different contribution of them {  \citep[see, e.g.,][]{Christlein2005, Guo2009, Iovino2010, Mercurio2010, Peng2010}} and, simultaneously, to understanding how they are related \citep[e.g.,][]{DeLucia2012}. 

At low redshift, the key environmental trends worth to be recalled are the following. Color and morphological fractions are very different functions of environment \citep{Bamford2009}. Morphology shows density dependence \citep{Bamford2009, VanDerWel2010}, as first found by \cite{Dressler1980a} considering only densities within clusters, and extended afterwards to the group regime  by \cite{Postman1984}. Trends are weaker but still present at fixed stellar mass \citep{Kauffmann2004, Blanton2005, Christlein2005, Weinmann2009, Calvi2012}. The red/passive fraction at fixed stellar mass is elevated in groups and clusters when compared to the general field \citep{Weinmann2006, Baldry2006, Kimm2009}. 
Galaxy groups seem to have on average early type galaxy fractions and passive fractions between those of clusters and those of the field \citep{Hansen2009, Kimm2009}. Therefore, since $\sim40\%$ of the galaxies found today in massive clusters were accreted from groups more massive than 10$^{13} {\rm M_\odot}$ \citep{McGee2009}, environmental effects rely heavily on what occurs in galaxy groups \citep[pre-processing,][]{Wilman2005, Wilman2009}.

Overall, the aforementioned trends continue to higher redshift (at least to $z\sim$1), but the difference between the dense environments and the field becomes smaller \citep{Gerke2007, Balogh2009, Kovac2010}.

Theoretical understanding of the possible mechanisms responsible for the observed environmental correlations is critical. Numerous processes have been proposed \citep[and references therein]{Boselli2006}. The most important are:
\begin{enumerate}
\item Strong interactions between galaxies and mergers. These are  efficient when the relative velocity among galaxies is low, therefore in groups of galaxies rather than in clusters \citep{Toomre1977}. Indeed, semi-analytic models  attempting to quantify the galaxy merger rate find that mergers do not occur frequently enough in massive groups and clusters to explain environmental correlations \citep{Bundy2009}. Given the high fraction of chance superpositions, the actual galaxy merger rates are difficult to determine in observations \citep{VanDokkum1999, Lin2010}. 
\item Harassment. This term refers to the cumulative effect of the gravitational interactions in high speed encounters that truncates the galaxy star formation in $\sim10^7$ years \citep{Moore1996}. Repeated heating events 
may cause a morphological transformation from a spiral into a spheroid \citep{Moore1998}. 
\item Strangulation. This is the removal of the galaxy hot gas supply that happens when the galaxy falls in a more massive dark-matter halo \citep{Larson1980, Cole2000, Balogh2000, Kawata2008}. This process inhibits the star formation, which dies out on a timescale of few $10^9$ yr \citep{Peng2015}. 
\item Gas stripping. The interaction between galaxies and the intergalactic medium produces a stripping of the gas that is mostly due to the effect of the ram pressure that is efficient in very dense environments, and when the relative velocity of galaxies is high. These conditions are achieved in the core of rich galaxy clusters \citep{Gunn1972} and typically quench the star formation in a few 10$^7$ yr. 
\end{enumerate}
In groups, the ram pressure is expected to be lower than in clusters. However, dwarf galaxies have lower masses than large spirals and therefore lower restoring pressures, and dwarf galaxies may be stripped of most of their gas.
Indeed, simulations of dwarf galaxies experiencing ram-pressure winds in groups showed that gas stripping can occur \citep{Marcolini2003, Hester2006}. In addition, large spirals have restoring pressures that decrease with radius and may be stripped of their outer disks.  Simulations by \cite{Roediger2005} showed indeed that spirals are stripped of some of their outer gas even in groups. 

In a small number of low-mass groups there is observational evidence for individual galaxies that have been ram-pressure stripped of their gas \citep[e.g.,][]{Bureau2002, Rasmussen2006, McConnachie2007, Sengupta2007, Freeland2011, Brown2017}.  Some of these groups also have X-ray detections of a hot intergalactic medium (IGM), consistent with the ram pressure stripping explanation. All the previous work has been done focusing on the neutral hydrogen (H{\sc i}). 

To date, there is no optical evidence for stripping in groups, like there has been in clusters via the discovery of the so-called ``jellyfish'' galaxies {  \citep[e.g.][]{Cortese2007, Smith2010, Ebeling2014}}.

In this context, we started the  survey GASP (GAs Stripping Phenomena in galaxies with MUSE),\footnote{\url{http://web.oapd.inaf.it/gasp}} an ESO Large program aimed at studying gas removal processes in nearby galaxies in different environments. GASP was granted 120 hr of observing time with the integral-field spectrograph MUSE mounted at the VLT and observations were concluded in April 2018. The program targeted 94 candidate stripping galaxies selected from the \cite{Poggianti2016} atlas, which was built on a systematic search for galaxies with signatures of one-sided debris or tails reminiscent of gas-stripping processes in optical images of clusters from WINGS \citep{Fasano2006} and OmegaWINGS \citep{Gullieuszik2015} surveys and groups from the Padova Millennium Galaxy and Group Catalog \citep[PM2GC,][]{Calvi2011}. Additional 30 between undisturbed and passive galaxies were observed as control sample. A complete description of the survey strategy, data reduction, and analysis procedures is presented in \cite{Poggianti2017a}(Paper I). First results on single  objects in clusters are discussed in \citealt[(Paper II)]{Bellhouse2017}; \citealt[(Paper III)]{Fritz2017}; \citealt[(Paper IV)]{Gullieuszik2017}; \citealt[(Paper V)]{Moretti2018}; and in lower-density environments in \citealt[(Paper VIII)]{Vulcani2017c}; \citealt[(Paper VII)]{Vulcani2018}.

 In this paper we present the analysis of four GASP galaxies that are part of the same group. 
The group is not virialised yet - therefore the mass estimate is quite uncertain - and is  forming at the intersection of two filaments. This peculiar environment gives us the possibility to {  characterise} the impact of the {  low-density environments} 
on the properties of the gas and stars of their members. Indeed, galaxies falling in along filaments connecting group and clusters are expected to have a very different infall experience from those accreted through largely empty regions \citep{Bahe2013}. 
{  We stress that this work characterises the spatially resolved propertied of galaxies belonging to the same low mass system. In the literature, many works have focused on stripped galaxies \citep[see, e.g.,][to cite a few]{Merluzzi2013, Fumagalli2014, Fossati2016, Merluzzi2016, Consolandi2017} in clusters, but none of them targeted the low halo mass regime and carefully characterise more than one or two stripped objects in the same system.}

In what follows, we will first present the MUSE data observations and analysis (Sec. \ref{sec:data}) for the four GASP galaxies. Then, we will characterise the group hosting these galaxies as a whole (Sec. \ref{sec:env}) and the integrated properties of its members (Sec. \ref{sec:prop}). Then, we will focus on the resolved properties of the GASP galaxies (Sec. \ref{sec:results}) and relate them to the position of the galaxy in space and velocity. We will emphasize the differences among the galaxies and discuss whether or not they feel the effect of the environment their are embedded in. Finally, we will discuss possible scenarios that describe the evolution of the objects and test whether ram pressure stripping can be invoked to describe the observations (Sec. \ref{sec:disc}).  We will conclude drawing the final conclusions (Sec. \ref{sec:conc}).

\begin{figure*}
\centering
\includegraphics[scale=0.4]{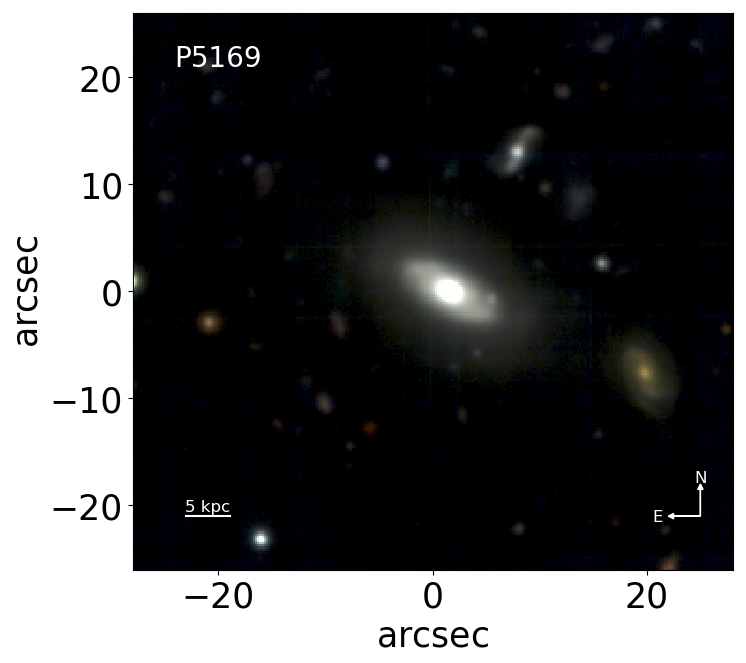}
\includegraphics[scale=0.4]{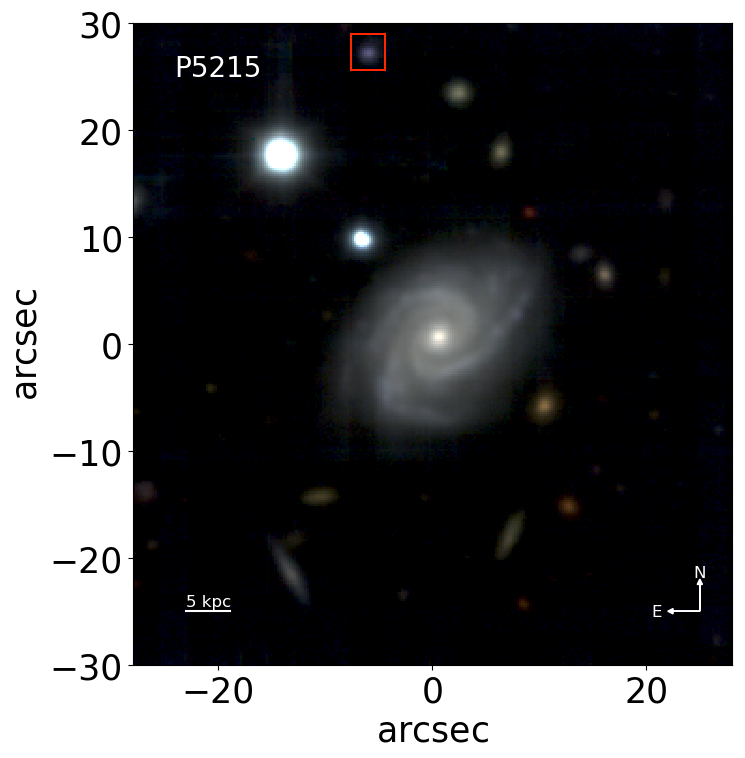}
\includegraphics[scale=0.4]{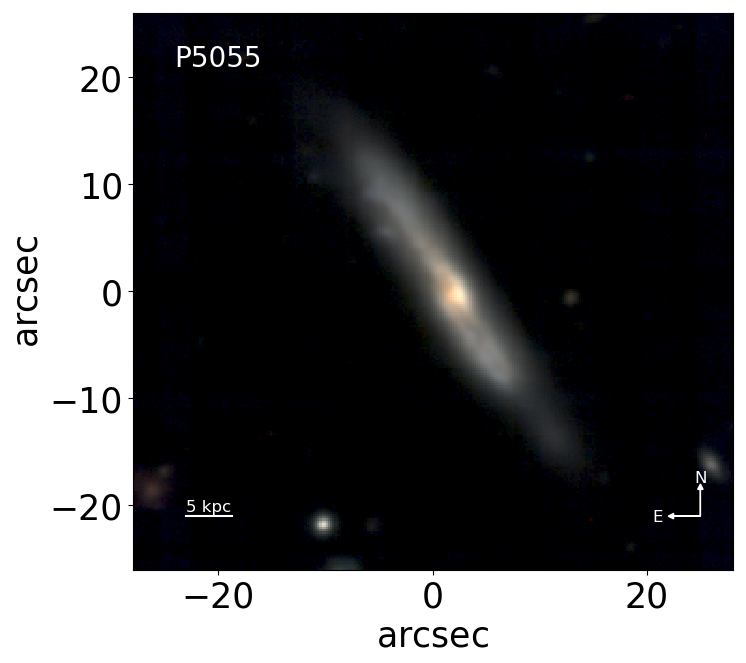}
\includegraphics[scale=0.4]{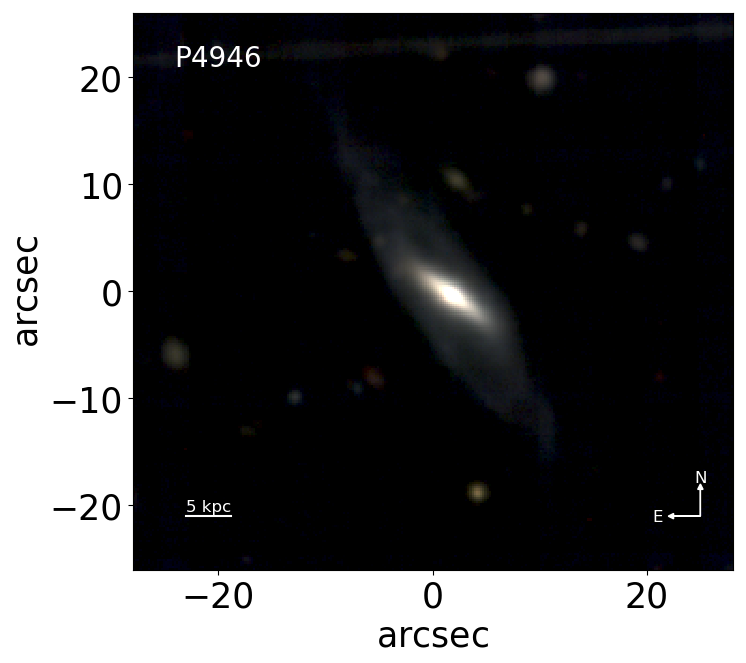}
\caption{RGB images of the targets. From  left to right, top to bottom: P5169, P5215, P5055 and P4946 are shown. The reconstructed $g$, $r$, $i$  filters from the MUSE cube have been used. 
North is up, and east is left.  At the galaxy redshift, 1$^{\prime\prime}$ = 1.233 kpc, see scale. In P5215, the red box shows the cloud that will be further analysed in Sec. \ref{sec:disc_P5215}. \label{fig:rgb_image} }
\end{figure*}

We adopt a \cite{Chabrier2003} initial mass function (IMF) in the mass range 0.1-100 M$_{\odot}$. The cosmological constants assumed are $\Omega_m=0.3$, $\Omega_{\Lambda}=0.7$ and H$_0=70$ km s$^{-1}$ Mpc$^{-1}$. This gives a scale of 1.223 kpc/$^{\prime\prime}$ at the group redshift, which is $z = 0.06359$. 

\section{GASP data} \label{sec:data}
\subsection{The targets}
The four GASP galaxies analysed in this paper are presented in Figure \ref{fig:rgb_image}, which shows the color composite images of the four targets, obtained combining the reconstructed $g$, $r$ and $i$ filters from the MUSE datacube. 

Three of the galaxies here analysed (P5215, P5055 and P4946) have been selected because they showed hints of morphological asymmetries in the B-band imaging that could be caused by unilateral forces such as ram-pressure stripping \citep{Poggianti2016}.\footnote{{  The B-band images can be downloaded from the online version of \citep{Poggianti2016}.}} {  Specifically, P5215 showed asymmetric spiral arms, with the ones toward South-East much more bended than the one towards North. P5055 showed a tail toward South West, and P4946 an extended halo with signs of stripping toward North.} P5169 instead was selected for being passive and served as control sample galaxy of the GASP project. All galaxies are drawn from the Millennium Galaxy Catalog \citep{Liske2003,Driver2005} and selected   from the PM2GC \citep{Calvi2011}. Morphologies have been assigned by \cite{Calvi2012}, {  using MORPHOT \citep{Fasano2012}, an automatic tool designed to reproduce the visual classifications.}

P5169 (upper left corner of Fig.\ref{fig:rgb_image}) has been classified  as an early-spiral galaxy (Sa); P5215 (upper right) is an almost face-on late spiral (Sc). Its spiral arms are very well defined.
P5055 (bottom left corner) is an almost edge-on late spiral galaxy (Sc).
Dust attenuation effects are evident from the image. 
Finally, P4946 (bottom right corner), {  classified by \cite{Calvi2012} as an S0 galaxy}, presents a quite concentrated body and  an extended disk characterised by a different orientation with respect to the bulge. The presence of a bar is detected. A ring that joins together the edges of the bar, leaving empty regions inside, is tentatively seen. 

In   Sec. \ref{sec:env} we will discuss in detail the integrated properties of the objects. 

\begin{table*}
\caption{Properties of the targets. For each galaxy, the ID, redshift, coordinates, total stellar mass, rest frame (B-V) color, morphology, and distance from the group center, in unit of $\rm R_{200}$  are given. \label{tab:gals}}
\centering
\begin{tabular}{lrrrrrrrrrrr}
\hline
  \multicolumn{1}{c}{ID} &
  \multicolumn{1}{c}{z} &
  \multicolumn{1}{c}{RA} &
  \multicolumn{1}{c}{DEC} &
  \multicolumn{1}{c}{$\log({\rm M_\ast}/{\rm M_\odot})$}  &
  \multicolumn{1}{c}{(B-V)$_{r-f}$}  &
  \multicolumn{1}{c}{SFR}  &
  \multicolumn{1}{c}{T$_M$} &
  \multicolumn{1}{c}{d$_{\rm r 200}$}  
  \\
  \multicolumn{1}{c}{} &
  \multicolumn{1}{c}{} &
  \multicolumn{1}{c}{(J2000)} &
  \multicolumn{1}{c}{(J2000)} &
  \multicolumn{1}{c}{}&
  \multicolumn{1}{c}{(mag)}&
  \multicolumn{1}{c}{(${\rm {\rm M_\odot} \, yr^{-1}}$)}&
  \multicolumn{1}{c}{} &
  \multicolumn{1}{c}{} \\
\hline
P5169 & 0.06355& 10:18:13.79 &+00:03:56.591 & 10.47 &0.718& 0&Sa&1.182\\
P5215 & 0.06308 & 10:16:58.24 & -00:14:52.876 & 10.46 & 0.587 & 3.9  & Sc& 2.368  \\
P5055 & 0.06095& 10:18:08.54 &-00:05:03.141 &10.24& 0.935&1.0 &Sc&0.346\\
P4946 & 0.06217 & 10:18:30.81 & +00:05:05.001 & 10.98 &0.778 & 1.0 & S0 &1.608\\
\hline\end{tabular}
\end{table*}

Table \ref{tab:gals} summarises some important information that will be obtained in the following and further used and discussed throughout the paper. 
It is interesting to note that all galaxies have similar mass (within a factor of 5), but very different star forming properties. 

\subsection{Observations and data reduction}
As all the other galaxies of the GASP project, the targets were observed in
service mode with the MUSE spectrograph, mounted at the Nasmyth focus of the UT4 VLT, at Cerro Paranal in Chile. All observations were taken in clear, dark-time,  <0$\farcs$9 seeing.
For each galaxy, a total of four 675 seconds exposures were taken with the Wide Field Mode. P5169 was observed on 2017, January 22; P5215 on 2016, December 29; P5055 on 2017, December 13; and P4946 on 2017, December 15. 

The data reduction process for all galaxies in the GASP survey is presented in  \citetalias{Poggianti2017a}.  

For all galaxies, we average filtered the datacubes in the spatial direction with a 5$\times$5 pixel kernel, corresponding to 1$^{\prime\prime}$ \citepalias[see][for details]{Poggianti2017a}. 

\subsection{Data analysis}\label{sec:analysis}
The procedures  used to analyse all galaxies of the GASP survey are extensively presented in \citetalias{Poggianti2017a}, so we refer to that paper for all the details. 
In brief, the  spatially resolved quantities obtained from the MUSE reduced datacube, corrected  for extinction due to our Galaxy, that can be used to characterise the targets are:

\begin{itemize}
\item total fluxes and kinematic properties of the gas, obtained exploiting the  \kube  \citep{Fossati2016} code;
\item kinematic properties of the stars, obtained using the  Penalized Pixel-Fitting (pPXF) software \citep{Cappellari2004}, which works in Voronoi binned regions of given S/N \citep{Cappellari2003} and smoothed using the two-dimensional local regression techniques (LOESS) as implemented in the Python code developed by M. Cappellari;\footnote{\url{http://www-astro.physics.ox.ac.uk/~mxc/software}}
\item properties of the stellar populations, such as star formation histories, luminosity and mass weighted ages, surface mass densities, obtained running the spectral fitting code  \sinopsis \citepalias{Fritz2017}; 
\item  dust extinction  A$_V$  from the absorption-corrected Balmer decrement.
\end{itemize}
In the following plots showing results for the gas, we will plot only the spaxels with \Ha S/N$> 3$ that have contiguous spaxels with similar S/N. 
In addition, for the spaxels in the outskirts of the galaxies and in possible isolated clouds, 
we will  plot only the spaxels whose velocity is within $3\sigma$ of the mean velocity of the portion of the galaxy close to each cloud. 
This approach helps to remove possibly spurious signal.

\section{The environment}\label{sec:env}
Before proceeding with the analysis of the GASP galaxies, we   present the analysis of the host system. 

We exploit the publicly available catalog of groups and clusters based on the spectroscopic sample of  galaxies of the SDSS data release 10, flux-limited at  m$_r$= 17.77
\citep{Tempel2014_g}. 
Using a friends-of-friends method, \cite{Tempel2014_g} identified  28  galaxies that belong to the system of interest. They computed the velocity dispersion from the mean group velocity and redshift, following the standard approach, finding a value of 354 \kms.  They also estimated a virial radius of $\sim 500$ kpc and a halo mass of $\sim 10^{14.1} {\rm M_\odot}$, using a \cite{Nfw1997} profile. 
\begin{figure*}
\centering
\includegraphics[scale=0.25]{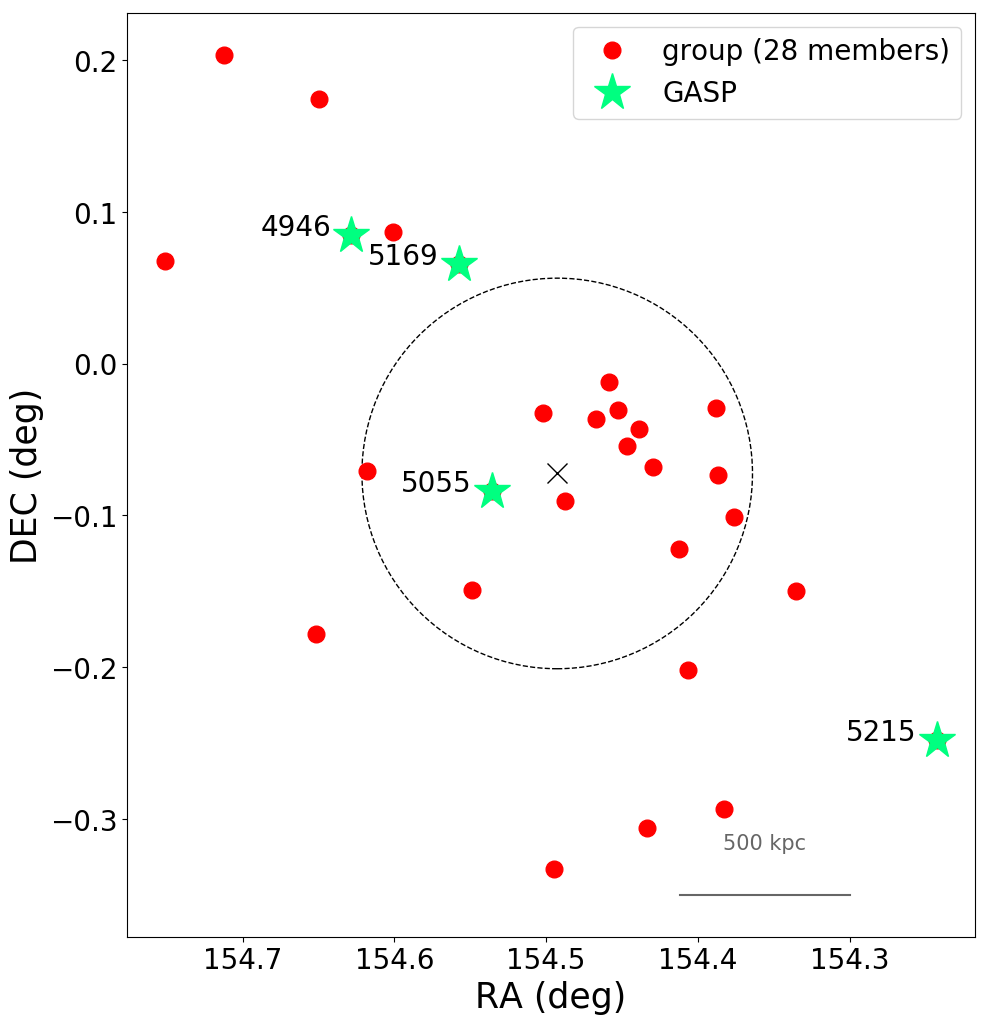}
\includegraphics[scale=0.25]{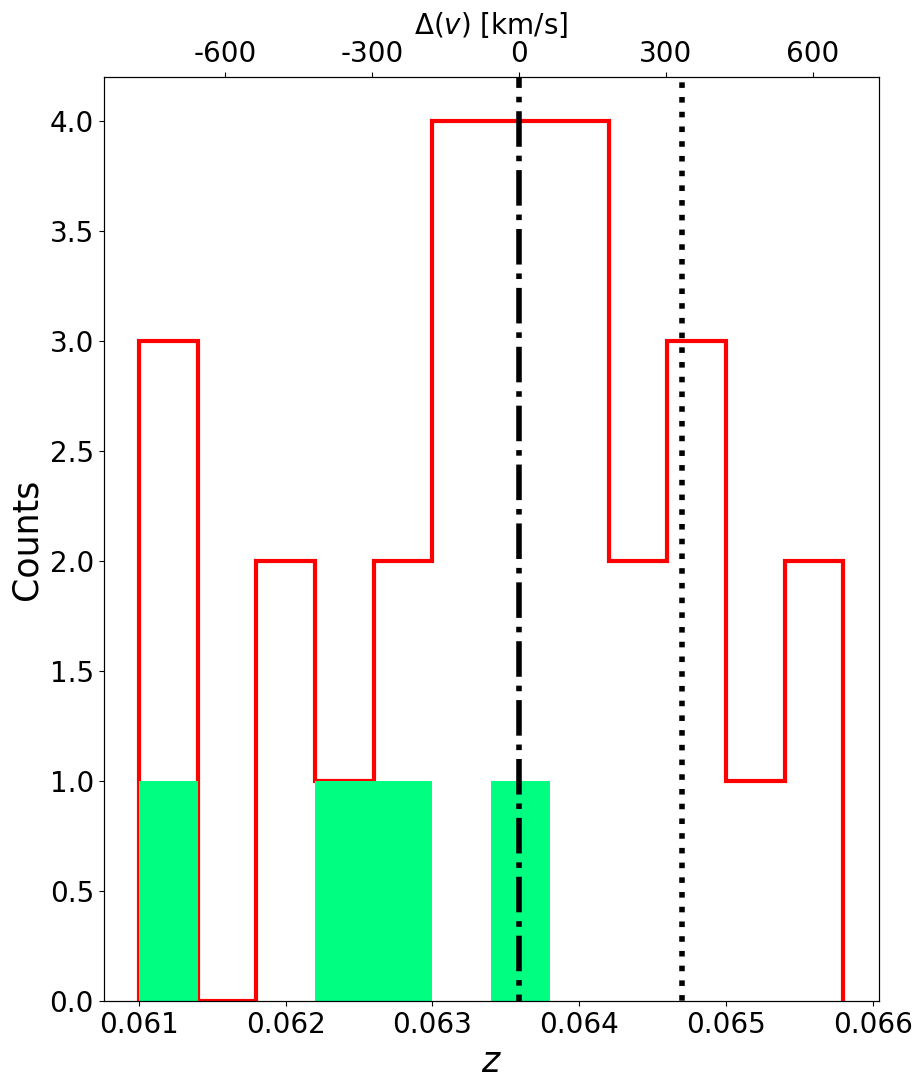}
\caption{Left: Position in the sky of the group and of its members. The galaxies that will be further analysed in the paper are highlighted with green stars.  The black dashed  circle indicates the virial radius of the group. The scale in the bottom right corner shows 500 kpc at the redshift of the group. Right: Redshift distribution of the galaxies in the group. GASP galaxies are highlighted in green. Dotted line is the group redshift given in \citet{Tempel2014_g}, dash-dotted line is the redshift adopted in this work. The upper axis shows the velocity difference with respect to the group redshift.
\label{fig:group} }
\end{figure*}
The left panel of Figure \ref{fig:group} shows the position in the sky of the group and of its members. The GASP galaxies that will be further analysed in the paper are highlighted with green stars. The spatial distribution of galaxies is unbalanced towards one side, suggesting that spherical symmetry has not been reached yet. {   Half of all group members are located outside the virial radius, suggesting that this group is not gravitationally bound yet.}
The halo mass estimate is therefore most likely an overestimation of the true halo mass. 
{  Unfortunately, no X-ray data are available for this system therefore we do not have an estimate of the density of the intra-group medium, a key ingredient to obtain reliable estimates of the halo mass and to characterise the processes affecting its members.}

{  Estimating the halo mass from the velocity dispersion using the relation given e.g. in \cite{Poggianti2006}, we obtain a value of $\sim 5.3\times 10^{13} {\rm M_\odot}$. This might be a more reliable value, though still an upper limit given the uncertainties on the estimate of the velocity dispersion.}

\begin{figure*}
\centering
\includegraphics[scale=0.5]{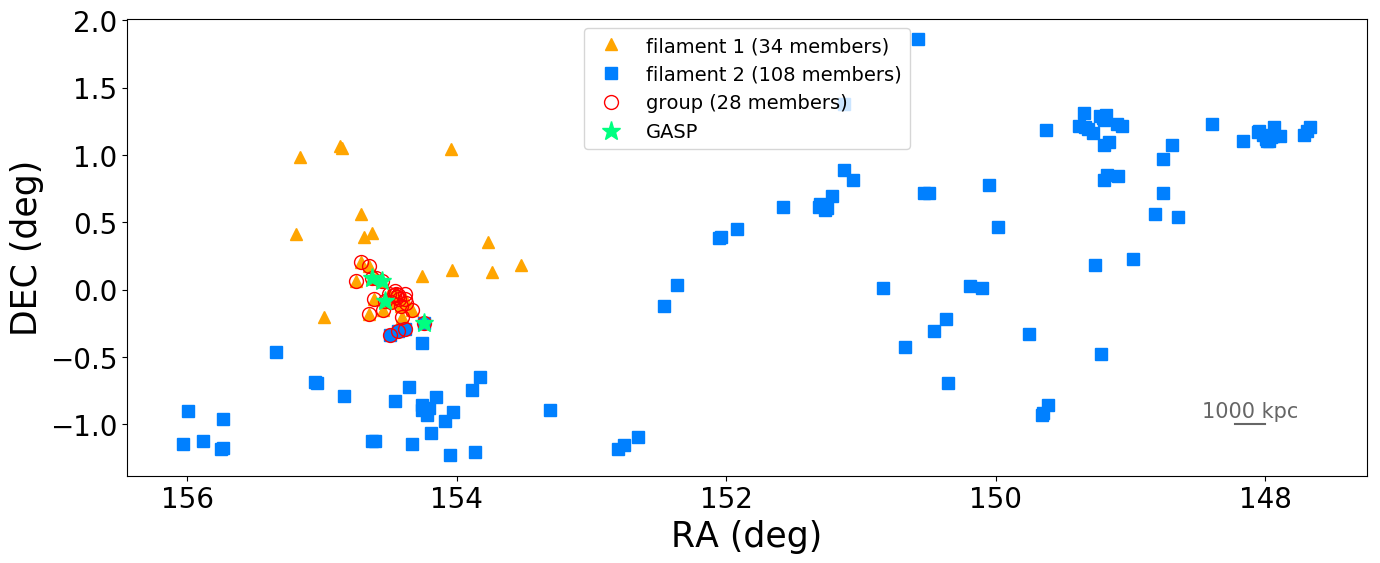}
\includegraphics[scale=0.5]{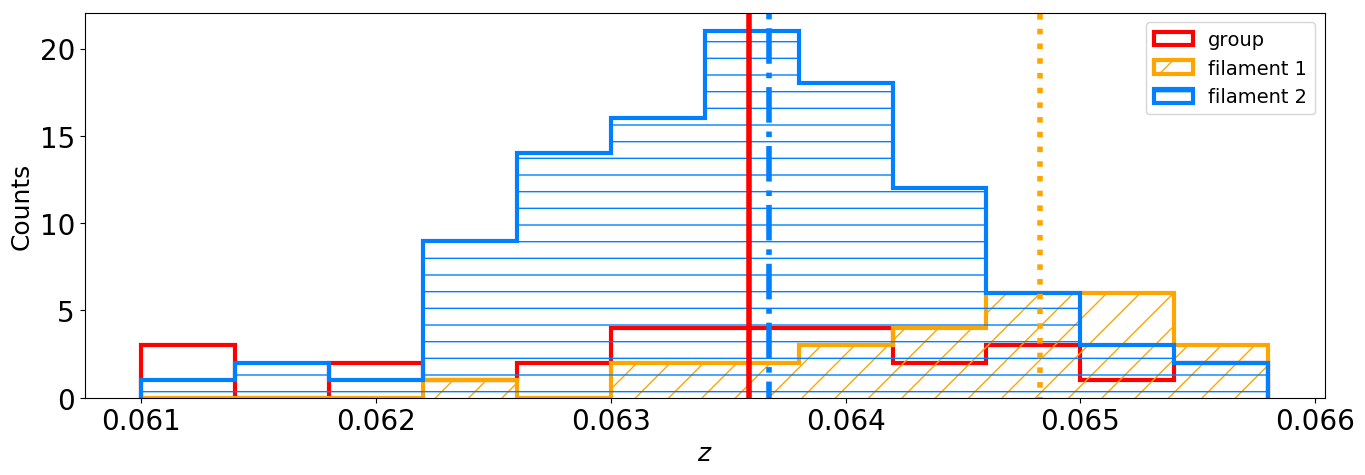}
\caption{Top: Position in the sky of the group \citep[from][]{Tempel2014_g} and of its members, along with the filaments \citep[from][]{Tempel2014_f} flowing onto the group. The galaxies that will be further analysed in the paper are highlighted with green stars. The scale in the bottom right corner shows 1000 kpc at the redshift of the group. {  Bottom: Redshift distribution of the galaxies in the group (red) and in the filaments (1: orange, 2: light blue). Vertical dotted lines represent the redshift of the systems.} \label{fig:filament} }
\end{figure*}

The morphology of the structure might be due to the fact that the group is located at the intersection of two filaments, as shown with different symbols in Fig. \ref{fig:filament}. The definition of the filaments is taken from \cite{Tempel2014_f}  who based their analysis on the same sample of the galaxies of SDSS data release 10 used by \cite{Tempel2014_g}. The first filament includes 34 members and extends towards North; {  its members have a quite flat redshift distribution, whose median value is 0.06483, corresponding to a velocity difference of $\sim$ 370 \kms with respect to the group.} The second filament extends mainly towards East for more than 3\degree\ in the sky. It includes 108 galaxies {  whose redshift distribution and median redshift are similar to those of the group}. As a result, galaxies in the group tend to lie along a stripe, going from South-West to North-East. 

 \citet{Tempel2014_g} computed the virial radius, but, since the structure is not relaxed, this quantity has to be taken with caution when used to characterise the extent of the structure and the  virialised region. Of the GASP galaxies,  P5055 is the  closest to the group center, while  the others lie beyond the computed virial radius. The redshift distribution  (right panel in Fig. \ref{fig:group}) is quite symmetric, but the  redshift provided by \citet{Tempel2014_g} seems not to be representative of the group. We therefore recompute the median redshift of the system, finding a value of  0.06359. This value will be used in the forthcoming analysis.

Table \ref{tab:groups} presents some useful information regarding the group. 
All values are drawn from \cite{Tempel2014_g} with the exception of the redshifts, 
as explained above.
\begin{table}
\caption{Properties of the group hosting  the galaxies. Except for z, which we computed, values are taken from \citet{Tempel2014_g}. The redshift (z$_{\rm gr}$), the coordinates (RA$_{\rm gr}$ and DEC$_{\rm gr}$), the number of group members (N$_{\rm gals, \, gr}$), the velocity dispersion ($\sigma_{\rm gr}$), and the virial radius R$_{\rm vir, \, gr}$ 
are given.   \label{tab:groups}}
\centering
\setlength{\tabcolsep}{2pt}
\begin{tabular}{crrrrrrrrrrrrrr}
\hline
  \multicolumn{1}{c}{z$_{\rm gr}$} &
  \multicolumn{1}{c}{RA$_{\rm gr}$} &
  \multicolumn{1}{c}{DEC$_{\rm gr}$} &
  \multicolumn{1}{c}{N$_{\rm gals, \, gr}$} &
  \multicolumn{1}{c}{$\sigma_{\rm gr}$} & 
  \multicolumn{1}{c}{R$_{\rm vir, \, gr}$}    \\ 
  \multicolumn{1}{c}{} &
  \multicolumn{1}{c}{(J2000)} &
  \multicolumn{1}{c}{(J2000)} &
  \multicolumn{1}{c}{} &
  \multicolumn{1}{c}{(\kms)} &
  \multicolumn{1}{c}{(kpc)} \\ 
\hline
  0.06359
    &154.49272	&-0.07237	&28	& 354 & 576 \\ 
\hline
\end{tabular} 
\end{table}
This group is clearly a structure in formation and it is therefore plausible to expect signs of distortions on the properties of its members.

\section{Global properties of the group members}\label{sec:prop}
Using the pieces of information collected in the PM2GC catalog, we can study the integrated properties of the group members, and then compare them with galaxies in other PM2GC groups, to investigate if they are peculiar or  similar to the general field population. 24/28 galaxies of the group are present in the PM2GC sample. 
Note that besides the galaxies discussed  in this work, in \cite{Poggianti2016} there are no other galaxies members of this group showing signs of distortion in the B-band images suggesting  gas removal processes in action, nor tidal interactions.  
For this comparison, for consistency with all the other group galaxies for which we have no GASP data, we will use stellar masses obtained from the photometry  following the \cite{BdJ2001} approach \citep{Calvi2012} and SFR obtained from running \sinopsis on the fiber spectrum of the galaxy \citep{Poggianti2013a}. As a consequence, the adopted values here are different from those given in Tab.\ref{tab:gals} that were obtained by integrating all the MUSE spaxels belonging to the galaxies and then running \sinopsis. {   It is well known that in general different methods adopted to compute stellar masses give values that can differ up to a factor of 2-3 \citep[see, e.g.][]{Kannappan2007, Pozzetti2007, Gunawardhana2011}. This is due to the fact that each approach includes different assumptions on the choice of e.g. the IMF, the star formation history, the stellar population models that inevitably impact the results. Our values are indeed concordant within the expected factor.}

\begin{figure}
\centering
\includegraphics[scale=0.4]{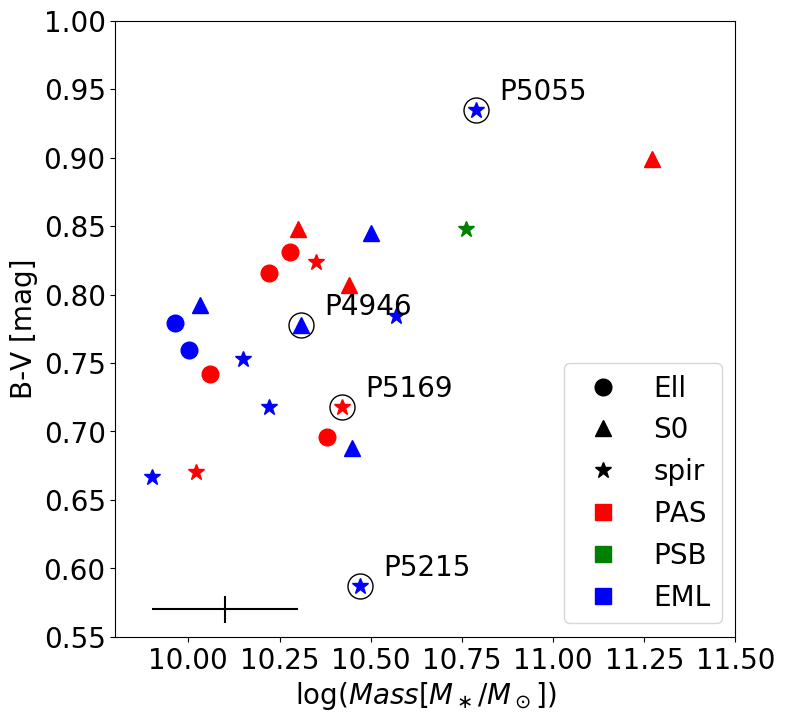}
\caption{Rest frame (B-V) Color- stellar mass diagram for galaxies belonging to the group. {  The median error bar is shown in the bottom left corner.} Galaxies with different morphological type (Ellipticals, S0s, spirals) are indicated with different symbols, galaxies of different spectral type (Passive, Post Starburst, Emission line) are indicated with different colors, as described in the caption. GASP galaxies are highlighted with an empty circle.  
\label{fig:cmd} }
\end{figure}

Figure \ref{fig:cmd} shows the  rest-frame (B-V) vs. stellar mass plane for the group members, distinguishing galaxies with different morphological and spectral types. Morphologies  are drawn from \cite{Calvi2012}, spectral types from \cite{Paccagnella2018}. 6 galaxies are classified as ellipticals, 8 as S0s, 10 as spirals. 12 galaxies are emission line galaxies, 1 is a post starburst, 10 are passive. Therefore, there is not a predominant type. 

All galaxies present quite red (B-V) colors (median $\sim 0.8$), the bluest galaxy is P5215, while the reddest is P5055. The other two GASP galaxies have intermediate colors. Note that a typical separation between blue and red galaxies is around (B-V)$\sim$ 0.6-0.7 \citep{Valentinuzzi2009}.

Overall, there is no 1:1 correspondence between morphology, spectral type and color: out of the 10 spiral galaxies, only 6 are emission line galaxies and only one has blue colors. On the other hand, 2/6 ellipticals show signs of emission. Three GASP galaxies are emission line objects, one is passive; as already mentioned 3 are spirals, one is an S0.

Besides the expected color-stellar mass relation, no other trends as a function of stellar mass are visible: elliptical galaxies are also found among the least massive systems of the group, emission line galaxies among the most massive ones.

From this analysis, we can infer that, even though being still in formation, this group has already undergone/is still undergoing a number of processes that are affecting the properties of their galaxies. 
The fact that we are observing galaxies whose color does not correspond to what is expected based on their morphology  suggests that the physical processes in action must operate on  different timescales on the star formation activity (color and spectral type) and structural (morphology) properties.   
 
\begin{figure*}
\centering
\includegraphics[scale=0.45]{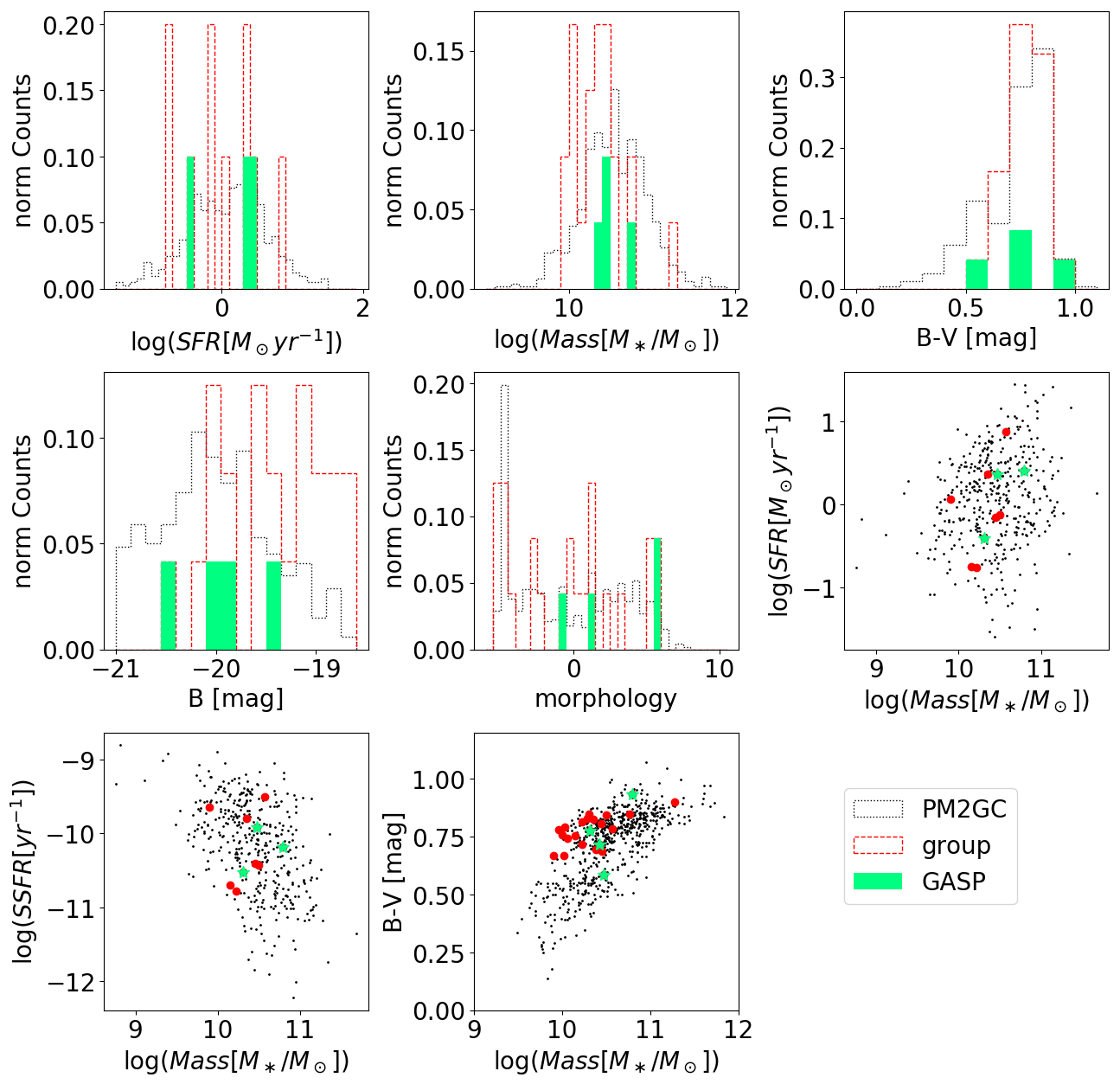}
\caption{Properties of the galaxies included in the group compared to those of all PM2GC galaxies, applying a cut  at $m_r<17.7$. Gasp galaxies are highlighted with red colors. 
\label{fig:properties} }
\end{figure*}

Figure \ref{fig:properties} compares the properties of the group to the properties of all the groups in the PM2GC sample at $0.04<z<0.1$ adopting the same magnitude cut as \cite{Tempel2014_g}, $m_r<17.7$. 
Overall, the galaxies in the group are systematically redder, fainter and less massive  than those in other groups, they therefore lie on the upper envelope of the color mass diagram of the groups.  However, they lie on the typical SFR- and SSFR-Mass relation (see also \citealt{Poggianti2016} and  Calvi et al. submitted) and their morphological distribution is {  statistically} similar. 
The Kolmogorov-Smirnov test is able to firmly detect differences  when stellar masses {  ($p_{value}=$0.02), magnitudes ($p_{value}=10^{-3}$) and colors ($p_{value}=$0.2) are compared.}

The top-middle and middle-left  panels of Fig. \ref{fig:properties} also show a relatively big mass/magnitude gap in the analysed group: {  the mass difference between the most massive galaxy and the second most massive one is 0.5 dex, the magnitude difference between the brightest and second brightest galaxy is 0.9 mag}. The difference in the masses/magnitudes of the two most massive/brightest galaxies has been widely used as an optical parameter related to mass assembly of groups and clusters. Systems of galaxies where most of the mass has been assembled very early, develop a larger magnitude gap compared to systems that form later \citep[e.g.,][]{Ponman1994, Khosroshahi2004, Khosroshahi2007,  VandenBosch2007, Barnes1989, DOnghia2005}. This piece of evidence therefore may suggest this group is somewhat relaxed, {  despite its spatial distribution}.

If we compare the group properties to those of the other PM2GC groups with similar number of members, we find that it is still redder than the others (plot not shown). This result is most likely due to the fact that the group is forming at the intersection of two filaments, therefore galaxies falling in have been already preprocessed in the filaments \citep[see, e.g.,][]{Poudel2017}.

The analysed group is therefore a peculiar environment and it seems to host galaxies with a variety of properties. {  This group therefore might not be representative of the general group population in the local universe. }
As mentioned in Sec.\ref{sec:intro}, many processes might be responsible for influencing galaxy properties. 

In the next section we will  present the analysis of 4 GASP galaxies and look for signs of possibly different physical mechanisms in action. 

\section{SPATIALLY RESOLVED PROPERTIES OF THE GASP GALAXIES}\label{sec:results}
\begin{figure*}
\centering
\includegraphics[scale=0.26]{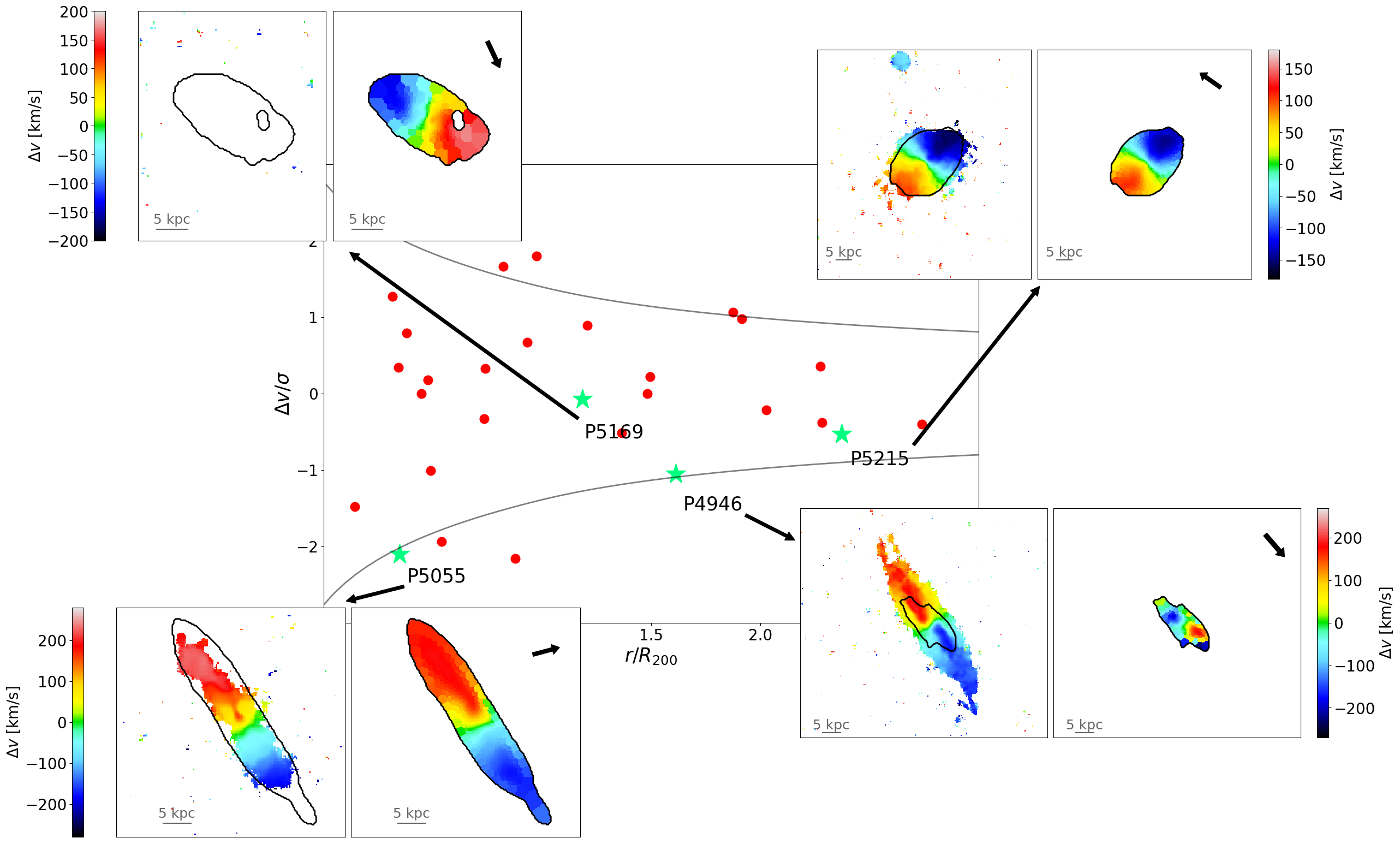}
\caption{Projected phase space diagram for the galaxies in the group. The four GASP galaxies are highlighted with green stars. The gray curve corresponds to the 3D (un-projected) escape velocity in a NFW halo with concentration {  c = 4} for reference. {  This value is appropriate for low mass systems \citep{Biviano2017}}. For each GASP galaxy, in correspondence to its position on the phase space diagram, the ionised gas velocity field and stellar velocity field are shown, using the same color scale. Black contours show the maximum extent of the stellar disk, as identified by pPXF. The black arrows in the upper right corners show the direction of the group center.
\label{fig:pps} }
\end{figure*}

We are now in the position of characterising the spatially resolved properties of the four GASP galaxies. 
Figure \ref{fig:pps} presents the stellar and ionised gas velocity fields for each galaxy linked to its location in the so-called phase space diagram for the group. This plot shows the projected  distance\footnote{{  This diagram is limited by uncertainties caused by the possible presence of sub-structures (in un-virialized systems), and projection effects which makes the measured distances and velocities lower limits to the real values, as discussed in detail in  \citetalias{Jaffe2018}.}} from the group center plotted against the line-of-sight velocity relative to the group velocity, normalized by the group size and velocity dispersion, respectively. We adopt as group size the virial radius measured by \cite{Tempel2014_g}, which is the most reliable estimate at our disposal. 
{  Note that, even though both the group velocity and the size of the virial radius as poorly defined for this group, they are  the same for all galaxies, therefore different values would simply produce a rigid shift to the points.} This diagram is a  tool for improving our understanding of environmental effects acting in groups and clusters of galaxies. Indeed, galaxies at different positions in their orbit occupy semi-distinct regions in projected phase- space diagrams \citep{Oman2013}. Different regions of phase-space could be associated with different amounts of tidal mass loss \citep{Smith2015} 
and/or different strength of ram pressure stripping \citep[Paper IX]{Jaffe2015,Jaffe2018}.

Even though the galaxy group is not relaxed, its members  are distributed inside the typical trumpet-shaped region delimited by the escape velocity in relaxed systems. Cosmological simulations \citep{Rhee2017, Haines2015} show that in these diagrams, galaxies that have been in a cluster for a very long time have lower velocities and clustercentric radii, because they have had enough time to sink into the potential well of the cluster. On the other hand, infalling galaxies approach the cluster core with high relative velocities at all clustercentric distances. Observations of galaxies in clusters confirm that, due to ram-pressure stripping by the ICM, gas rich galaxies avoid the ``virialised" region of phase-space \citep{Jaffe2015, Yoon2017}, which is populated mostly by passive galaxies quenched by the hostile cluster environment \citep{Jaffe2016, Paccagnella2017}. Observations of H{\sc i} in group galaxies have further suggested that the processes quenching galaxies in clusters could scale to the group regime \citep[see results from][]{Catinella2013, Hess2013, Jaffe2012, Jaffe2016}.

\subsubsection*{P5169}
\begin{figure}
\centering
\includegraphics[scale=0.2]{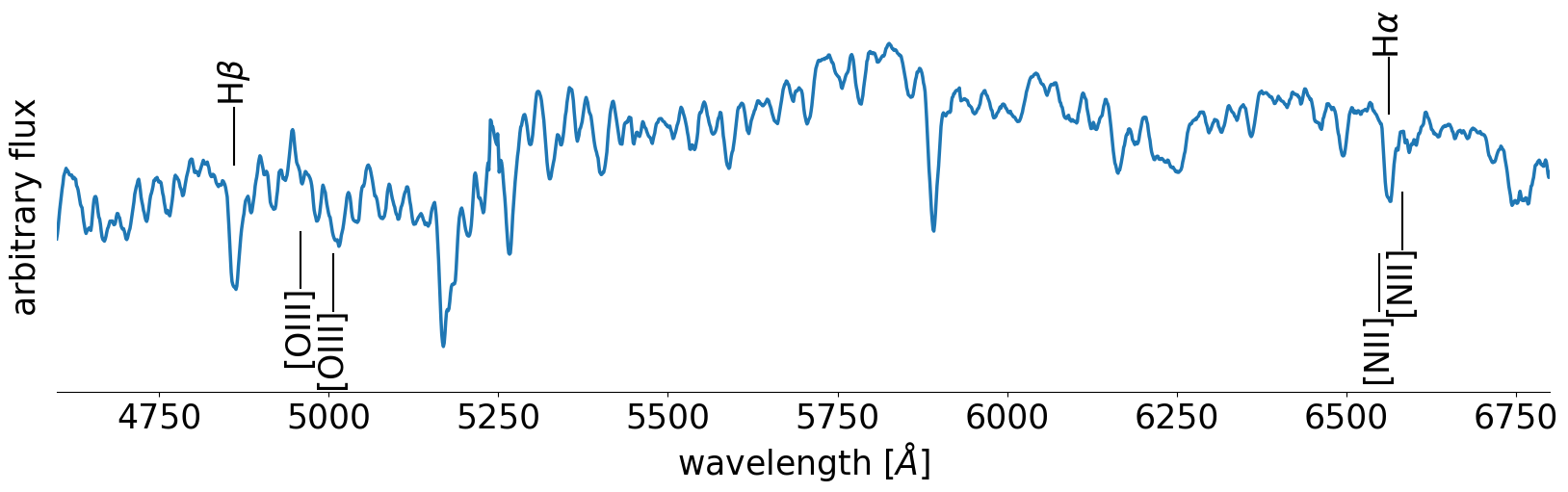}
\caption{P5169. Integrated spectrum over a radius of 5 arcsec. \label{fig:P5169_extra} }
 \end{figure}
 
Focusing on the galaxies of interest, P5169 has very low projected velocity ($\Delta(v)/\sigma$ close to 0), consistent with that of the oldest group members. 
The combination of its low velocity with its spatial location at $\sim1-1.5\times R_{200}$ from the group centre however is inconclusive, as in that region of phase-space there is a large mix galaxies with different times since infall \citep{Rhee2017}. 
P5169 has  no ionised gas, and has a spectrum typical of a passive galaxy (both its total integrated spectrum and that obtained integrating over a radius of 5 arcsec; see Fig.\ref{fig:P5169_extra}), which supports the scenario that it has been in the group for a long time. 

The stellar velocity field is regular, spanning a range of -190$<$v (\kms) $<$190, with an uncertainty of 15 \kms. Uncertainties on the stellar motion are the formal errors of the fit calculated using the original noise spectrum datacube and have been normalized by the $\chi^2$ of the fit. Note that a background galaxy has been masked while measuring the kinematics of the galaxy.

\subsubsection*{P5215}
Moving our attention to the next galaxy, P5215 is the farthest galaxy and it occupies approximately the region where recently infalling galaxies  might be found. 
 P5215 has a regular stellar kinematics, indicative of a regular rotation, ranging from $v=-150$ \kms to $v=110$ \kms, with a median error of 29 \kms. The median velocity dispersion is 30 \kms. The velocity field of the ionised gas unveils few interesting features. The rotation in the main galaxy body follows the stellar rotation, with a velocity of -170$< v$ (\kms) $<$170 and a median error of 4 \kms, and its extent is slightly larger in the south-west and north-east parts, and in the south-eastern direction several detached clouds are visible.  We remind the reader that we plot only reliable spaxels, having removed those that might be affected by spurious signal. We are therefore confident that these observed clouds are real. In addition, their velocity  is the one expected given their position with respect to the galaxy, suggesting that they indeed belong to the object. To further assess the values obtained for the velocity of the clouds, we integrated the spaxels of each cloud and run  \kube on the integrated spectra. Velocities obtained on the spatially resolved and integrated spectra are  in agreement.
The gas velocity dispersion of this galaxy is overall quite low ($\sim15$ \kms, plot not shown), indicative of a dynamically cold medium, except in the very center where it reaches values of 50 \kms. The median error is 4 \kms.
Another peculiarity of the ionised gas velocity field of  P5215 is the presence of a large cloud at 27.5$^{\prime\prime}$ north of the galaxy. This object has a diameter of $\sim 5^{\prime\prime}$, a velocity of $\sim -50$\kms, ranging from -75 <v (\kms) < -40, a gas velocity dispersion of $\sim 25$ \kms  and has no significant stellar counterpart. Indeed pPXF has not been able to detect any signal in correspondence of the position of the object. This object will be further discussed in Sec. \ref{sec:disc_P5215}. 

Using  the emission lines located within the cube observed range (i.e. H$\beta$, [OIII] 5007 \AA{}, [OI] 6300 \AA{}, \Ha, [NII] 6583 \AA{}, and [SII] 6716+6731 \AA{}), we  build  diagnostic diagrams \citep[BPT,][]{Baldwin1981}  to investigate the origin of the gas ionization and distinguish between regions photoionised by hot stars and regions ionised by shocks, LINERs and AGN. The lines' intensities are measured after subtraction of the continuum, exploiting the pure stellar emission best fit model provided by \sinopsis, to take into account any possible contamination from stellar photospheric absorption. Only spaxels with a $S/N> 3$ in all the emission lines involved are considered. These  diagrams indicate that the main source of ionization is star formation (plot not shown).

\subsubsection*{P5055}
\begin{figure*}
\centering
\includegraphics[scale=0.35]{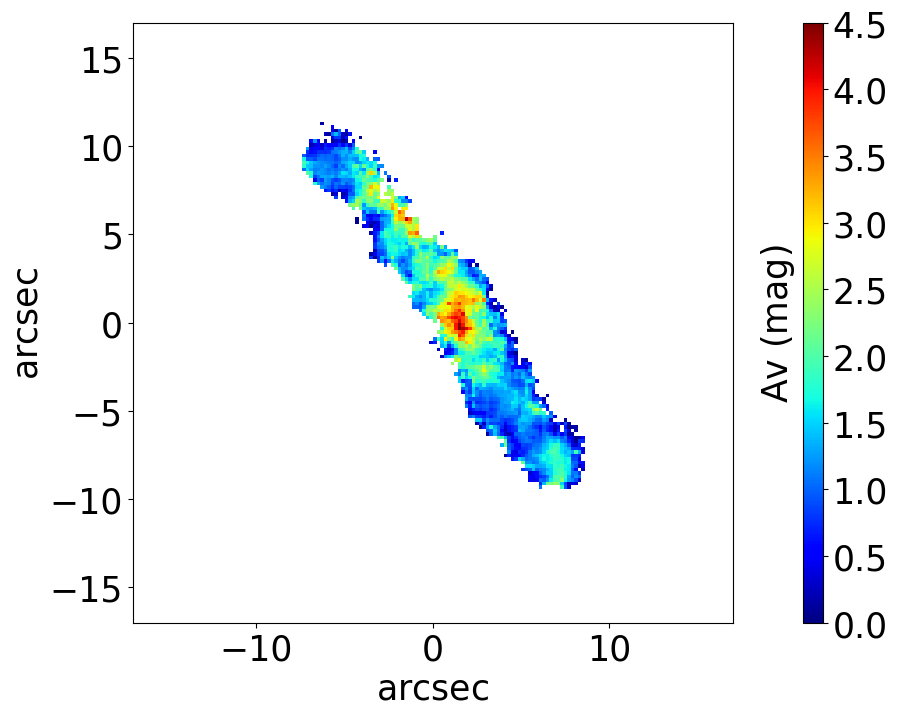}
\includegraphics[scale=0.35]{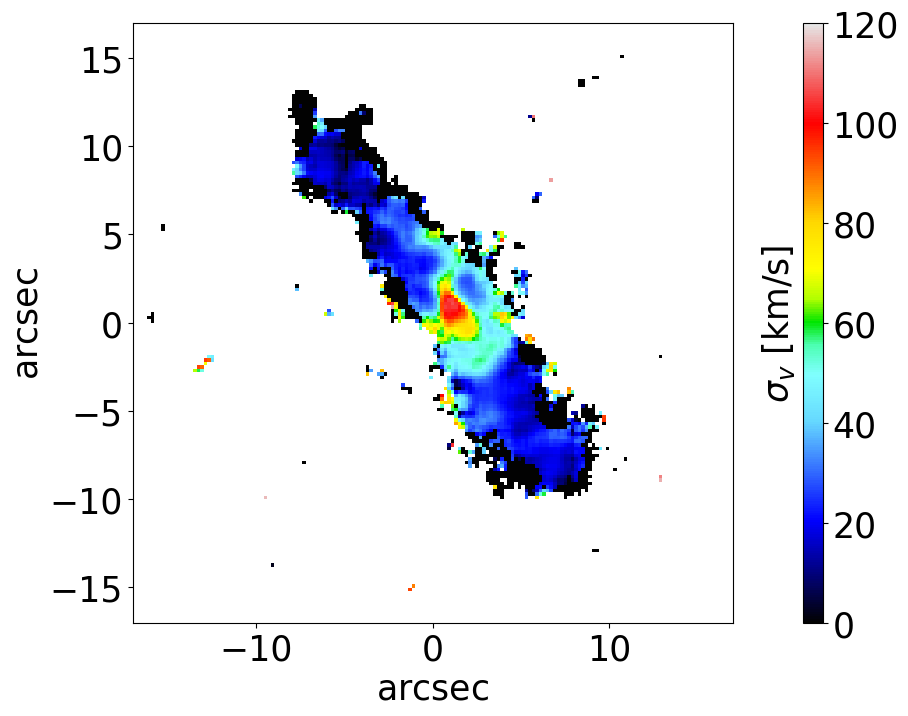}
\caption{P5055. Left: $A_V$ map. Only spaxels with a S/N(\Ha)> 3  are shown. Right: \Ha velocity  dispersion  map.  Only significant spaxels (see Sec. \ref{sec:analysis} for details) are shown.
 \label{fig:P5055_extra} }
 \end{figure*}

P5055 is not only the closest galaxy to the group center, but also the galaxy with the highest velocity difference with respect to the group's 
mean velocity.
This galaxy is in the region where 
ram pressure 
is expected to be 
highest (\citealt{Gunn1972}, \citetalias{Jaffe2018}). Its stellar velocity ranges between -180$<$v (\kms) $<$190, with a median error of 19 \kms.  
The North-East side is receding and the South West side approaching with respect to the observer. The horn visible in the southern part is probably due to a spiral arm, visible also in Fig. \ref{fig:rgb_image}.  The median velocity dispersion (plot not shown) is $\sim 70\pm30$ \kms. The irregular pattern observed in the galaxy center is probably due to the high extinction of the dust. Indeed, the left panel of Fig. \ref{fig:P5055_extra} shows extremely high values of $A_V$, especially in the galaxy center ($A_V\sim 4.5$), and this explain the very red color of this galaxy in the color-mass diagram. 

The ionised gas qualitatively follows the stellar velocity field, but it is much less extended, suggestive of a truncated disc. It is also slightly bended, in the direction opposite to the group center, and spans a relatively larger velocity range: -190$<$v (\kms) $<$230. The median uncertainty on the gas velocity is of 2.5 \kms. The right panel of Fig. \ref{fig:P5055_extra} shows that the velocity dispersion is very high in the core of the galaxy ($>100$ \kms), and close to 0 in the outskirts. The median velocity dispersion is $27\pm5$ \kms. Relatively high values of the velocity dispersion in galaxy cores are usually associated with the presence of AGNs. The analysis of the diagnostic diagrams does not show evidence of AGN activity, and indicates that the main source of ionization is star formation, even though a portion of the galaxy is characterised  by a Composite/LINER spectrum (plot not shown).
Note that  \cite{Veron2006, Veron2010},  using the ratio of \Hb and [OIII], classified P5055 (identified as J101808.6-000503)  as Seyfert 2. 
This galaxy might actually be hosting  an highly obscured AGN, so that it would not be detected by optical line diagnostics, similar to what discussed by \citetalias{Fritz2017}. Unfortunately there are no deep Chandra archive images to detect the presence of a luminous though highly absorbed X-ray source.

\subsubsection*{P4946}

\begin{figure*}
\centering
\includegraphics[scale=0.28]{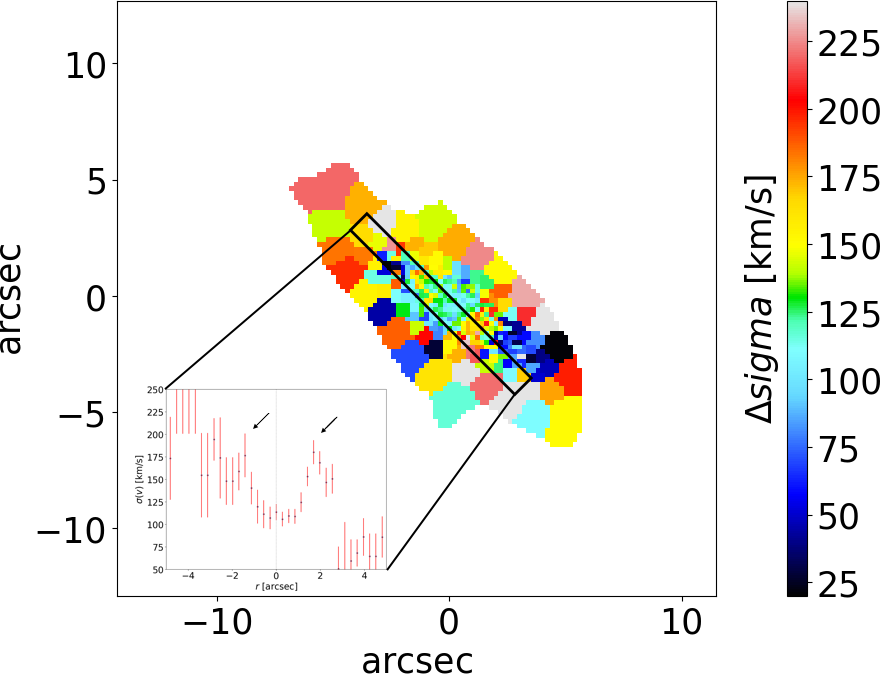}
\includegraphics[scale=0.35]{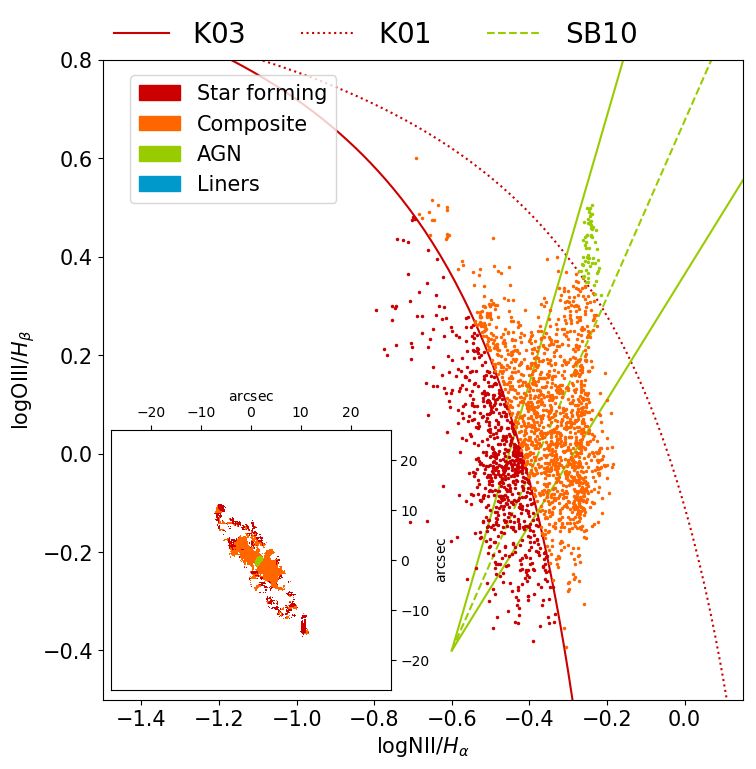}
\caption{P4946. Left: Stellar velocity dispersion map. The inset shows the stellar velocity dispersion profile in the inner region. {  The two arrows show the} double peak, indication of the presence of two counter-rotating stellar components. Right: BPT line-ratio  diagram for [OIII]5007/H$\beta$ vs [NII]6583/\Ha. Lines  are from \citet[][K03]{Kauffmann2003}, \citet[][K01]{Kewley2001} and \citet[][SB10]{Sharp2010} to separate Star-forming, Composite, AGN and LINERS. In the inset the BPT line-ratio map is shown. Only spaxels with a $S/N> 3$ in all the emission lines involved are shown. \label{fig:P4946_extra} }
 \end{figure*}

Finally, P4946 is located in the phase-space region where recently infallen galaxies might be found, similar to P5215. Its stellar kinematics ranges from -230 to 190 \kms, with a median error of 4 \kms. The kinematics are chaotic, a receding and approaching side of the galaxy are hardly identifiable. The left panel of Figure \ref{fig:P4946_extra} shows  the stellar velocity dispersion map, along with the profile extracted within a rectangular slit covering the central $3^{\prime\prime} \times 10^{\prime\prime}$  of P4946. Two off-centered maxima are evident, and are indication of two counter-rotating stellar components \citep[e.g.][]{Katkov2013}.

Focusing on the ionised gas, it is astonishing how much more extended than the stellar component it is, in both directions{  , as visible in Fig.\ref{fig:pps}}. It counter-rotates and is tilted with respect to  the stellar component. The velocity range is -200$< v$ (\kms) $<$190, with an uncertainty of 7 \kms. On both sides, semi-detached clouds are visible, probably related to the presence of spiral arms. These clouds have the same velocity as the galaxy, suggesting they follow the rotation of P4946 and belong to it. The gas velocity dispersion (plot not shown) is very high in the center, where the emission-line ratios are  consistent with gas being photoionised by the presence of an AGN, as shown in the right panel of Fig. \ref{fig:P4946_extra}. The same plot shows that the vast majority of the galaxy presents a composite spectrum, and only in the outskirts star formation is the dominant mechanism. This result is in agreement with the classification by \cite{Veron2006}, where P4946 corresponds to the galaxy J10183081+0005048.

\begin{figure*}
\centering
\includegraphics[scale=0.35]{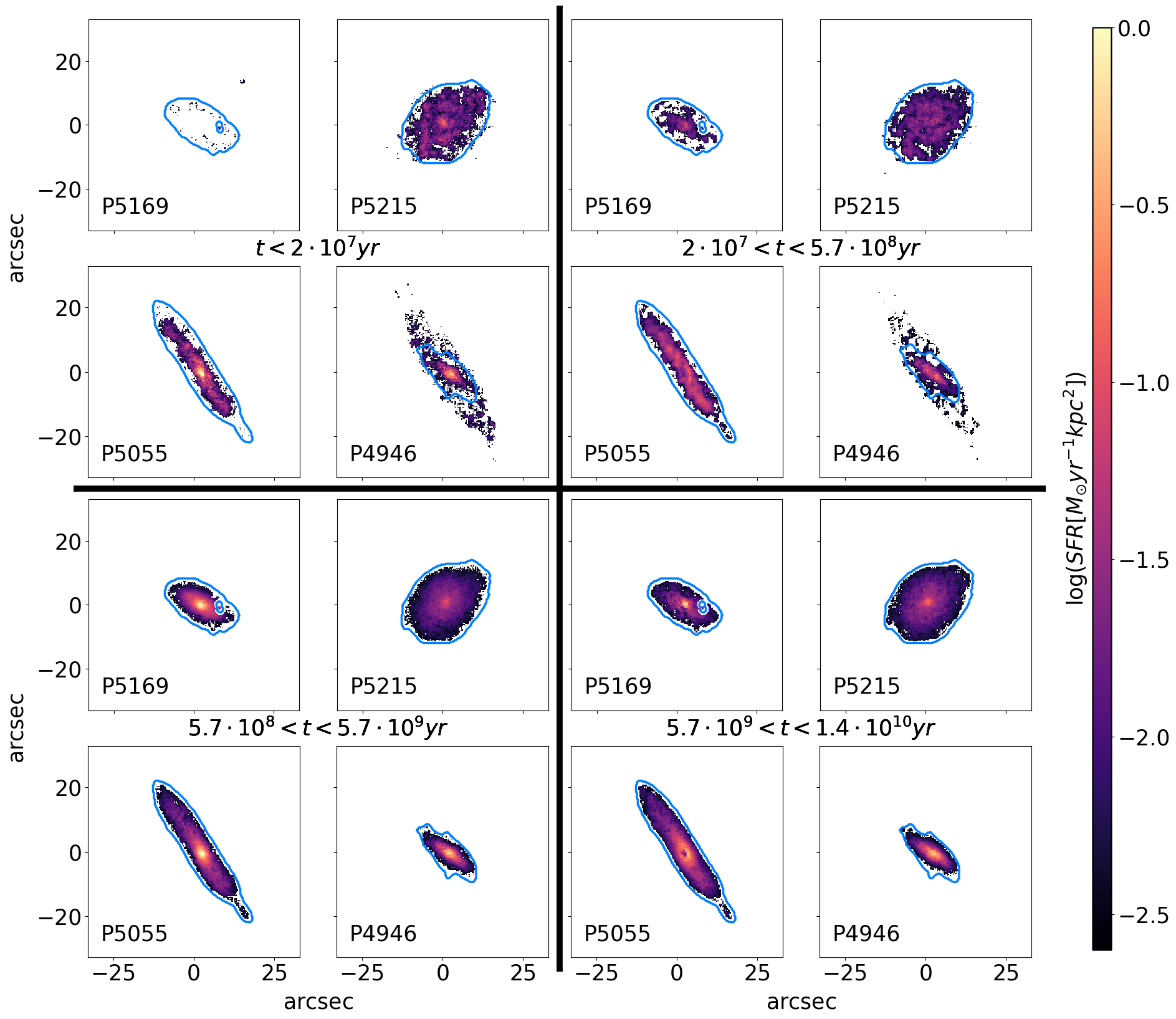}
\caption{Stellar maps of different ages, illustrating the average star formation rate per kpc$^2$ during the last $2\times 10^7$ yr (top left), between $2\times 10^7$yr and $5.7 \times 10^8$yr (top right), $5.7 \times 10^8$yr and $5.7 \times10^9$yr (bottom left) and $> 5.7 \times 10^9$yr ago (bottom right), for the four galaxies.  Light blue contours show the maximum extent of the stellar disk, as identified by pPXF (see Fig.\ref{fig:pps}). 
\label{fig:sfh} }
\end{figure*}

To summarize, three of the galaxies  present a quite regular stellar kinematics, indication that they have not undergone any merger event or suffered from tidal interactions that usually affect the orbits of the stars. Only P4946 is characterised by chaotic kinematics, that will be discussed later on.
The four galaxies present distinct gas distributions, suggesting that these galaxies are in different phases of their life and that their evolution might be driven by different physical processes. 

\subsubsection*{Star formation histories}
Figure \ref{fig:sfh} shows the galaxy star formation histories, that is the variation of the  SFR  with cosmic time.  We choose  four logarithmically spaced age bins in such a way that the differences between the spectral characteristics of the stellar populations are maximal (\citealt{Fritz2007} and \citetalias{Fritz2017}). 
Due to the code characteristics, when the spectra have a low signal-to-noise, \sinopsis tends to include an unnecessary small percentage of old (t>$5.7\times 10^8$) stars. To be conservative, we neglect the contribution of stars older than $5.7\times 10^8$ yr in low S/N spectra (S/N$<$3). The entire  contribution of young stars, instead, is taken into account, given the fact that it is estimated from the emission lines, which are more reliable features. The bottom right panels of Fig. \ref{fig:sfh} show the SFR that took place in the oldest age bin ($t>5.7\times 10^9$ yr ago), for the four galaxies respectively. During this epoch, the SFR is as extended as the stellar disk for P5215, P5055 and P5169, while in P4946 it is concentrated in the galaxy core.  In the following age bin ($5.7\times 10^8$ yr$<t<5.7\times 10^9$ yr), the  galaxy SFR and its extent do not change much for P5215, P5055 and P4946, while a possible small shrinking is visible for P5169. In the next  age bin ($2\times 10^7$ yr$<t<5.7\times 10^8$ yr), the SFR extent  considerably decreases for P5055 and P5169, and, to less extent, for P5215. In contrast,  it is much more extended for P4946  Finally, in the current SFR ($t<2\times 10^7$ yr), the SFR is still very high for P5215 and P4946, is negligible in P5169, and limited to the central regions of the galaxy for P5055. Note that the big cloud observed northern than P5215 appears only in the current SFR, but is not plotted here for scale reasons and will be shown later on. 

\section{Discussion}\label{sec:disc}
The galaxies presented in this work belong to the same group {  and infalling filaments} and provide us with {  the} possibility of investigating the role of the {  low-density} environment in shaping the spatially resolved properties of its members. In the previous section we have shown that these galaxies  are very distinct, suggesting that they are at different stages of their evolution and that are experiencing different effects from their host. In what follows, we summarize the most important results for each galaxy and present a picture describing their status. 

\subsection{Galaxy by galaxy interpretation}

\subsubsection{P5169}

P5169 is the only passive galaxy in this group that has been observed in the context of the GASP survey. It is at the edge of the virial radius of the group and its position on projected phase space suggests it has been part of the structure for a long time.

\begin{figure*}
\centering
\includegraphics[scale=0.3]{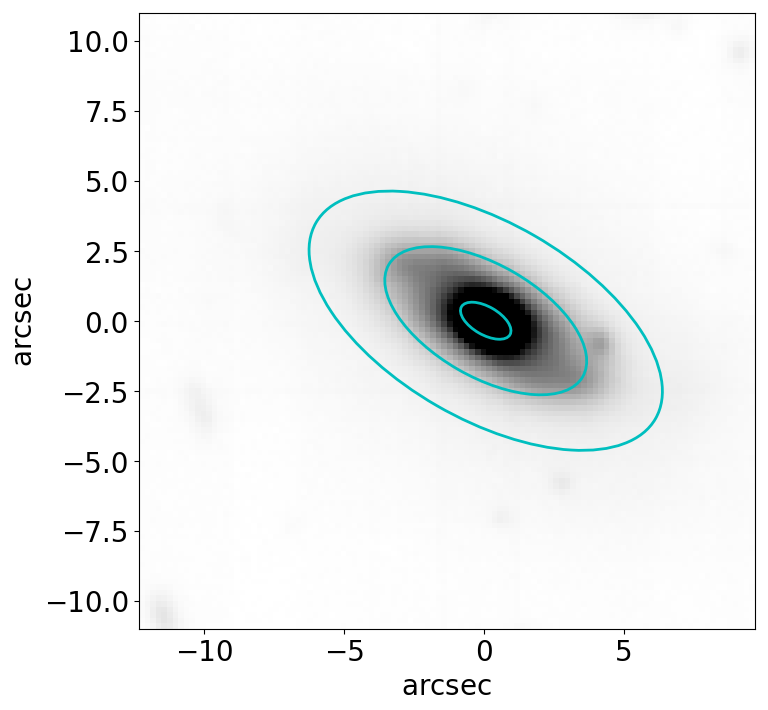}
\includegraphics[scale=0.4]{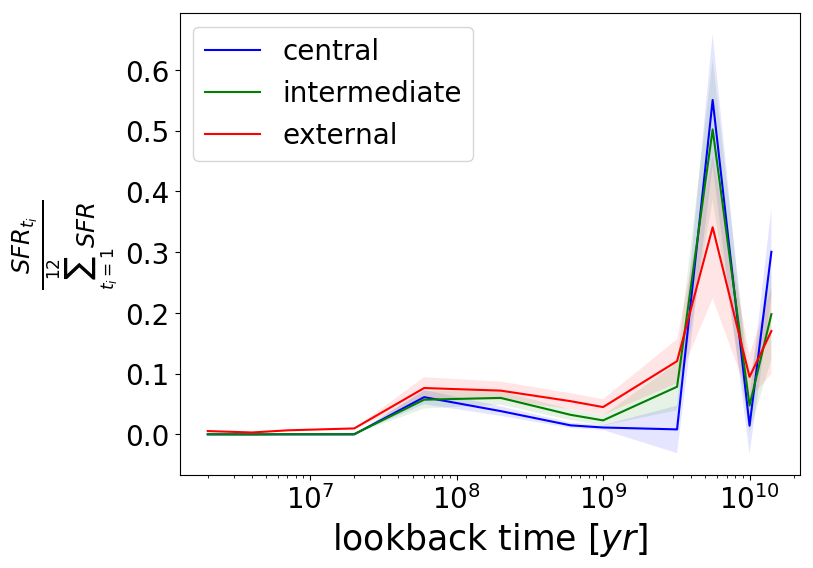}
\caption{P5169. Left: White light image with overplotted the annuli within which the star formation histories are computed. Right: Integrated Star Formation Histories in 12 age bins of the three regions highlighted in the left panel. Values are normalized so that the area below the curves is always equal to 1. {  Error bar are obtained propagating the formal errors to the fit.} \label{fig:P5169sfh} }
 \end{figure*}

Its integrated color and the analysis of the SFHs suggest that its star formation started declining $> 10^9$ yr ago and that the galaxy completely quenched only a few 10$^7$ yr ago. 
To better investigate how the quenching proceeded with time, Fig. \ref{fig:P5169sfh} shows the star formation histories in three different spatial annuli. To have a finer view of the quenching process, we plot here all the 12 age bins given by \sinopsis, without averaging in four bin as we did in Fig.\ref{fig:sfh}. We normalize the curve so that the area within each curve is equal to unity. This normalization allows us to compare the steepness of the star formation rate decline across time, rather than the absolute values, which are already visible in Fig.\ref{fig:sfh}. 

It emerges that in all the three different regions the star formation started declining between 3 to 5$\times 10^9$ yr ago. This 
epoch might correspond to the first infall of the galaxy onto the group. For the following $\sim 2\times 10^9$ yr, the galaxy has continued forming stars both in the central, intermediate and external regions, but at a systematically lower rate. The decline is very similar in the two outermost regions, while in the central part there are hints that $\sim 10^8$ yr ago an increase of star formation occurred, before the galaxy completely quenched $\sim 10^7$ yr ago in all regions.{  We note that this trend is unique among the galaxies presented in this paper: the other objects indeed show a peak of the star formation at $t<10^8$ yr (plots not shown.) } 

The shape of the star formation histories, characterised by an abrupt decrease, the observed time scales and the fact the star formation rate density maps do not show any asymmetry at any epoch are consistent with this galaxy exhausting its gas by strangulation. 
Our  analysis however  does not clarify what the strangulation mechanism is (e.g. hot halo environmental strangulation or via various preventive feedback mechanisms).

It is possible that the galaxy entered the group a few Gyr ago on a tangential orbit that never brought it close to the group center. P5169 has never strongly impacted with the ICM, therefore has never been affected by ram pressure stripping. Nonetheless, its gas reservoir has been truncated so the object has started consuming its remaining gas, until the exhaustion that occurred $\sim 10^7$ yr ago, when the galaxy became passive.  The fact that the galaxy just recently stopped forming stars is also in agreement with the position of the galaxy in the color-mass diagram (Fig. \ref{fig:cmd}): its intermediate color suggests that the galaxy is moving from the blue to the red sequence.

\subsubsection{P5215}\label{sec:disc_P5215}

\begin{figure*}
\centering
\includegraphics[scale=0.25]{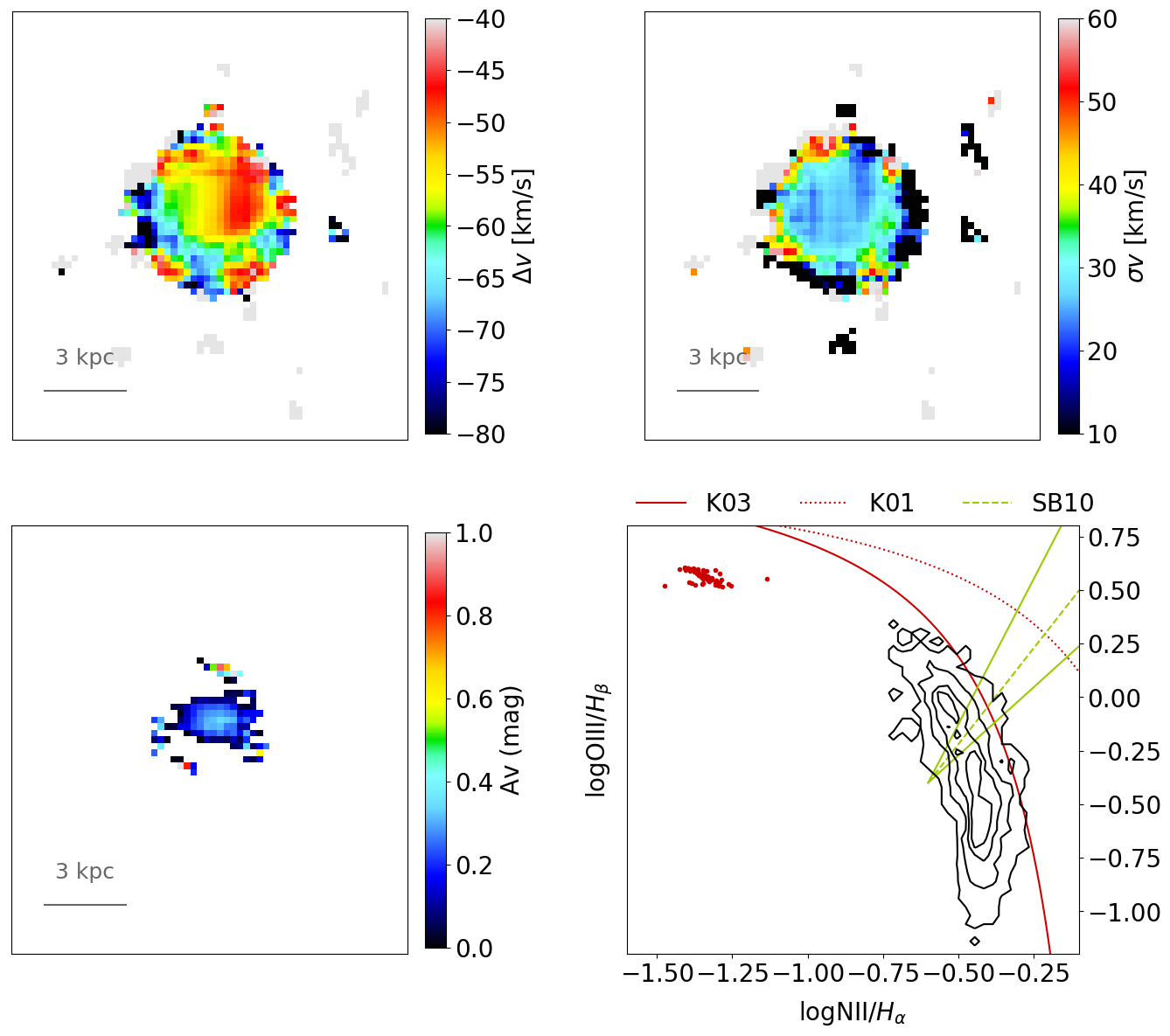}
\includegraphics[scale=0.3]{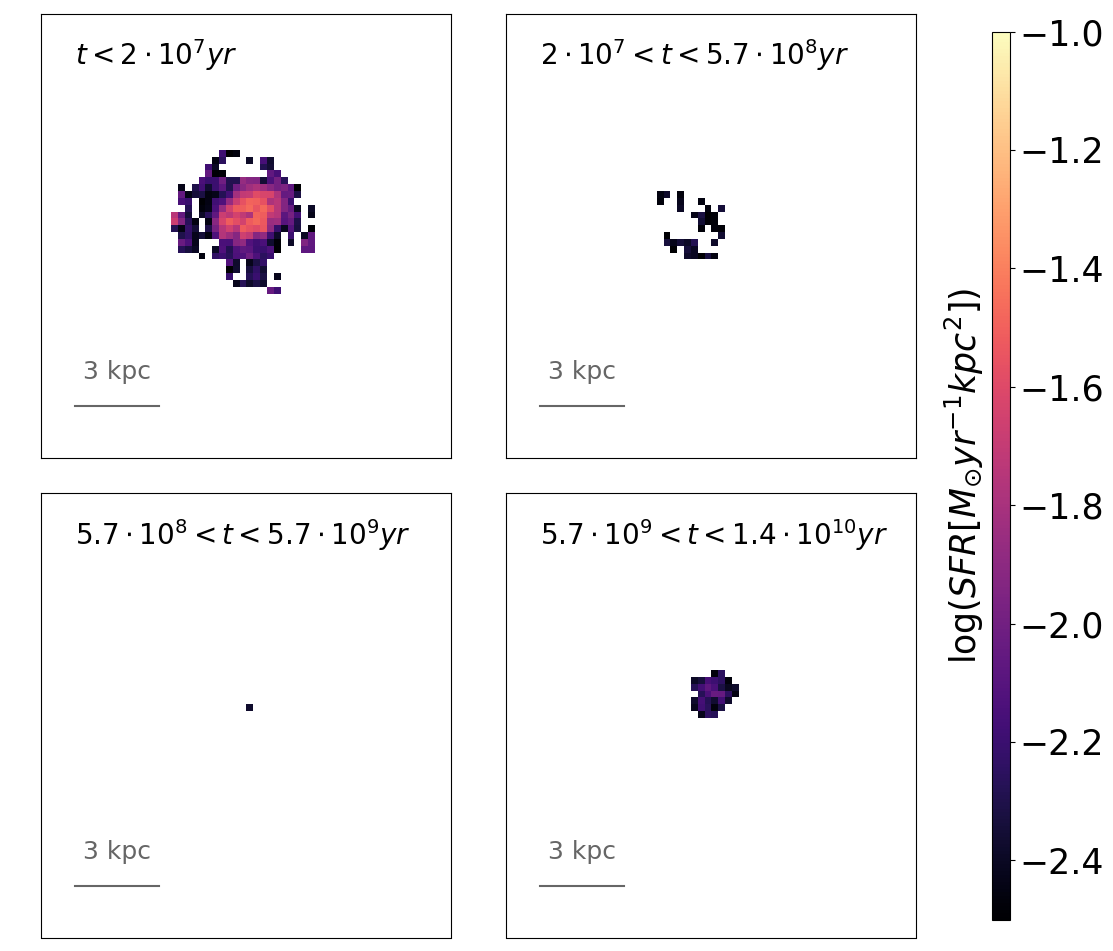}
\includegraphics[scale=0.4]{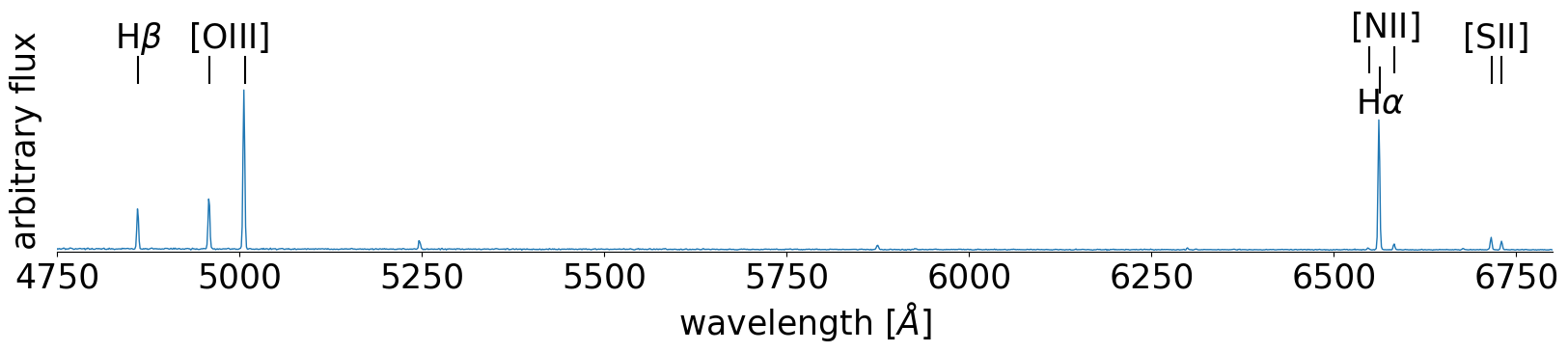}
\caption{Properties of the cloud located Northern to P5215 (upper panels) and its integrated spectrum (bottom panel). Ionised gas velocity and velocity dispersion (in the rest frame of P5215) dust maps, diagnostic diagram, and stellar maps of different ages. In the maps, the scale on the bottom right indicates 3 kpc. In the diagnostic diagram, the location of P5215 is shown as black contour plot. Lines represent regions with 1, 10, 30 and 50 points. 
\label{fig:P5215_cloud} }
\end{figure*}

P5215 is a face-on spiral galaxy located $\sim 2.4$ virial radii from the group center. It is farthest GASP galaxy from the group center and might be just about to fall into the group. Given its position in the projected phase space diagram (Fig.\ref{fig:pps}), it is expected to feel very little the influence of the group. Interestingly, P5215 is also embedded at the junction of two filaments, and it is at least qualitatively aligned to the bigger one. As discussed in Vulcani et al. (in prep., Paper XVI), filaments could also be effective in impacting the gas distribution of its members.  It is indeed very likely that the IGM density there is enhanced, rising  the possible ram pressure intensity. The galaxy could therefore be affected by stripping due to the presence of the filament \citep[see][for a theoretical approach]{Benitez2013}. As the galaxy is aligned to the filament and the velocity difference with respect to it must be low, the stripping, if any,  is expected to be relatively mild, and indeed overall, the main body of the galaxy is only marginally disturbed, being both the ionised gas and the stellar kinematics regular. The ionised gas is slightly more extended that the stellar disk.

Rather than stripping, numerical simulations by \cite{Liao2018} show that filaments can assist gas cooling and enhance star formation in their residing dark matter haloes. This scenario is consistent with the hypothesis that the densest regions in the circumgalactic gas get switched in their star formation when the galaxy impacts the low density IGM. In agreement with this, we indeed observe some detached clouds in the south-east part of the galaxy that seem to be switched on  behind the galaxy as it flows into the filament. In Paper XVI we  present other pieces of evidence for this  ``cosmic web enhancement''.

In P5215 we also observe 
 a $\sim 5^{\prime\prime}$ ionised gas cloud 27.5$^{\prime\prime}$ north of the galaxy. The position of the cloud is along the line connecting the galaxy and the group center. Figure \ref{fig:P5215_cloud} presents in detail some key properties of the cloud. Its ionised gaseous component  shows some rotation: the velocity difference between the two sides is $\sim 40$\kms. The object however does not follow the rotation of P5215, but counter rotates. The ionised gas velocity dispersion is almost constant across the object, except in the outskirts where the lower S/N prevents us from drawing solid conclusions. It is $\sim 30$\kms, indicative of a dynamically cold medium. The diagnostic diagram shows that the ionised gas is powered by star formation. The cloud has NII/\Ha values considerably lower than those of P5215, while OIII/\Hb values slightly higher.  Computing the ionised gas metallicity for both the cloud and the galaxy, using a modified version of the pyqz Python \citep{Dopita2013} v0.8.2 code (F. Vogt 2017, private communication), we obtain a value of  $12+\log[O/H]\sim 8.1$ for the cloud and of $\sim 9$ for the galaxy. Given their mass (see below), both objects lie on the typical metallicity-mass relation of local galaxies \citep{Tremonti2004}.

The $A_V$ of the cloud is 0.2-0.3 mag and reveals the presence of two possible spiral arms in formation. These are also detected in the map of the ongoing ($t<2\times 10^7$ yr) star formation rate. The latter is the only age bin where the star formation is non negligible and relatively extended. Indeed, already in the next age bin ($2\times 10^7<$t$<5.7\times 10^7$ yr) the star formation is consistently less extended and vigorous. For even older ages ($t>5.7\times 10^8$ yr) only a compact low star formation knot is seen. However, note that for $t>5.7\times 10^8$ yr the significance of the results is very low: if we apply a cut in S/N at 5, no signal is detected anymore.   These findings are therefore consistent with the result that the stellar continuum is very faint as no old stars are present (as also seen in the integrated spectrum). 
Integrating over the spaxels of the cloud, we obtain a stellar mass of 1.3$\times 10^8$ M$_\odot$ and a current SFR of 0.06 ${\rm M_\odot} \, yr^{-1}$. The latter has been obtained using the \Ha luminosity corrected for dust and stellar absorption and the \cite{Kennicutt1998a} relation. 
All the findings above suggest a recent formation for the cloud. 

In principle, the cloud could be a portion of gas ionised in the disk of P5215 and then stripped, but in this case it should travel at impossibly large speeds to reach large galactocentric distances before recombining and decaying. In fact, for a gas density $n$ = 10 $cm^{-3}$, the recombination time of hydrogen is about $10^4$ yr and once recombined the decay time is negligible \citep{Osterbrock2006}. Thus, the gas recombination lines  are visible only for a time shorter than this timescale after the gas is ionised. The recombination time goes linearly with the gas density. We can not estimate the gas density for this cloud following \cite{Proxauf2014}, as the  ratio of the [S II] 6716 and [S II]6732 lines is outside the interval for which the relation has been calibrated. However, given the measured ratio, we expect it to be lower than  $n = 10 \, cm^{-3}$. Assuming this value as an upper limit, and considering that the cloud is located at a distance from the disk of $\sim 35$ kpc, we obtain the gas should travel at most at a speed of $\sim 3.5\times 10^6$ \kms to get there in $10^4$ yr. This velocity is much higher than the velocities at which galaxies are typically moving within the IGM.

In addition, the fact that the gas of the cloud counter rotates with respect to the gas of P5215, and that the ionised gas metallicity is much lower suggest a different origin of the cloud from that of P5215.

We can also  exclude the hypothesis that the cloud is a low mass companion of P5215, as we do not detect any stellar continuum with old stars. 

We therefore conclude
that the cloud is most likely  a dense circumgalactic cloud switched on for compression as the galaxy flows into a filament, feeling the cosmic web enhancement.

\subsubsection{P5055}
P5055 is the galaxy that, according to the projected phase space, is in the best position to suffer from environmental processes, such as ram pressure stripping. Indeed, its ionised gas distribution is truncated with respect to the stellar disk, suggesting that some process already removed part of the existing gas, without affecting the star kinematics. 
The galaxy is very inclined and dusty, especially in the core, and this makes its integrated color red. 

Besides being very close to the group center, P5055 is also aligned to the filament it is embedded in. 

P5055 resembles JO36, a cluster galaxy in the GASP sample presented in \citetalias{Fritz2017}. Both galaxies are highly inclined, have a  truncated ionised gas disk, a disturbed gas kinematics,  present a horn in the southern region of the galaxy and are in a similar location on the phase-space diagram of the host structure. These pieces of evidence made \citetalias{Fritz2017} to conclude that in JO36 ram pressure was caught in a post-stripping phase. 
The moderately intense star formation of JO36 likely induced by shocks between the gas within the galaxy and the one in the ICM, consumed a substantial amount of gas. Their result is supported by numerical simulations: \cite{Kronberger2008} propose that loss of gas by ram pressure stripping, together with depletion due to star formation, is the reason for the decrease, and eventual quenching, of the star formation rate.

In Sec. \ref{sec:rp_modeling} we will perform some analytical calculations to test whether ram pressure can be invoked to explain the morphology of P5055.

\subsubsection{P4946}
P4946 is probably the most peculiar object of the group. It is an S0 galaxy composed by two counter rotating stellar disks. The galaxy is surrounded by an extended ($\sim 20$ kpc) ionised gas disk that formed stars only in the last $t<5.7 \times 10^8$ yr. It might be a polar ring, seen from an unfortunate angle. 

Galaxies with structural components with remarkably different kinematics are as rare as fascinating \citealt{Rubin1992, Ciri1995, Bertola1996, Galletta1996,Bertola1999, BertolaCorsini1999, Sarzi2000, Sarzi2001, Corsini2002, Corsini2012, Chung2012,
Pizzella2018}; \citetalias{Moretti2018}) and their origin is still poorly understood.

Extended counter-rotation of large stellar disks is usually considered as the signature of external processes as a past merging event or accretion \citep[e.g.,][]{Thakar1997, Puerari2001, Algorry2014, Bassett2017}.  Each of the two scenarios are expected to leave different signatures in the kinematics and stellar populations.
A major merger of two disk galaxies may end up with the formation of a disk galaxy with a massive counter-rotating disk under certain conditions \citep{Crocker2009}. 
Gas accretion with angular momentum opposite to that of the host galaxy can also produce a counter-rotating stellar disk. The newly-formed counter-rotating stellar disk is thinner and dynamically colder than the prograde disk. Stars formed in the secondary disk are expected to be younger and have similar metallicity to the gas from which it originated \citep{Algorry2014, Bassett2017}.

As we do not detect any signs of past merger in P4946, our  observations  point to the gas accretion hypothesis.

To properly characterise and date the different counter rotating components it would be necessary to analyse the {  Calcium triplet (CaT)} wavelength region \citep[see, e.g.,][]{Pizzella2018}, which allows to measure  differences in the stellar velocity dispersion of the two stellar components. At the redshift of P4946, the CaT is embedded in the sky lines and its analysis is beyond the scope of this paper.

The origin of the counter rotating disk could also be internal to P4946 itself. 
Indeed,   the galaxy  is also characterised by the presence of a bar. 
Barred galaxies host quasi-circular retrograde orbits, and the origin of the stellar counter-rotation observed in barred galaxies \citep{Bettoni1989, Bettoni1997} can be the result of internal dynamical processes \citep{Wozniak1997}.

Another characteristics of P4946 is the central AGN.
The combination of bar and AGN is not uncommon in local galaxies in groups: \cite{Alonso2014} found  that AGN spiral galaxies in groups are more likely to be barred than those in the field, indicating that galactic bars are stimulated in denser environments, and also that the physical mechanisms produced within groups and clusters of galaxies  might be related to the bar phenomenon \citep{Berentzen2007, Kormendy2012, Masters2012}. Barred AGN host galaxies also show an excess of population dominated by red colors, 
suggesting that bars produce an important effect on galaxy colors of AGN hosts \citep{Oh2012, Alonso2014}. 
As far as the properties of the host groups is concerned, at a given virial mass or total magnitude, the host groups of the barred AGN exhibit a larger fraction of red colors, than host groups of the corresponding unbarred AGN galaxies in the control sample 
\citep{Alonso2014}. 
The properties of P4946 and of its host group are in line with these findings: the group is particularly red as also P4946.

As an aside, we mention that we can  exclude that tidal interaction played a role: the closest neighbor, which is a group member, is a spiral, emission line galaxy at $\sim$120 kpc, 
5 times less massive 
than P4946 with a velocity difference of 525 \kms. According to \cite{Vollmer2005}, the acceleration $a_{tid}$ produced by this galaxy on the ISM of P4946 is much smaller than the acceleration from the potential of P4946, $a_{gal}$, making its effect negligible. 

To conclude, even though we can not detect any clear signs of some physical process typical of the group environment on this object, the gas availability that produced the counter rotating ionised gas disk is most likely external. 

\subsection{Modeling the ram pressure stripping intensity}\label{sec:rp_modeling}
In this section we aim at understanding whether ram pressure stripping can be invoked to explain the features observed in P5215 and P5055, which are the two group galaxies possibly showing signs of this mechanism.

The most direct way to infer the intensity of ram-pressure in a system is from X-ray observations that directly trace the density of the intra-cluster medium ($\rho_{ICM}$). This has been successfully done for massive clusters,  showing that $\rho_{ICM}$ increases with decreasing radius and, as a result, the ram-pressure ($P_{\rm ram}$) does too. The shape of the density profile however can vary from cluster to cluster, especially at the core. However, to give an idea, at a distance of $1\times R_{200}$ from the centre of a massive cluster ($M_{halo}\sim 10^{15} M_{\odot}$), ram pressure can have values around $P_{\rm ram}= 10^{-13}$~N~m$^2$ at a velocity equal to the velocity dispersion of the cluster \citepalias[see Fig. 3 in][for a comparison of $P_{\rm ram}$ in 2 clusters]{Jaffe2018}.
For low mass systems that are not X-ray bright, such as galaxy groups and filaments, it becomes difficult to measure the density of the intra-group medium ($\rho_{IGM}$) and thus the $P_{\rm ram}$.   
It is possible however to infer such information using  high-resolution cosmological hydrodynamic simulation. 
In particular, \cite{Bahe2013} has  computed the $\rho_{IGM}$, mean relative velocity, and resulting $P_{\rm ram}$ in clusters, groups, filaments and even voids. 
They focus on two halo mass bins. For massive clusters ($M_{\rm host}=1.6 \times 10^{15} M_{\odot}$) the estimated $P_{\rm ram}$ ranges from $\sim 3 \times 10^{-13}$ to $\sim 3 \times 10^{-15}$~N~m$^2$ in the region $1-3\times R_{200}$. These values drop to $\sim 3 \times 10^{-14}$ to $\sim 10^{-15}$ N~m$^2$ in the same region for lower mass groups ($M_{\rm host}=1$ to $3 \times 10^{13} M_{\odot}$) comparable to the galaxy group studied in this paper. 
However, \cite{Bahe2013} found that there is significant scatter in $P_{\rm ram}$ in the outer regions of clusters and groups, that can be explained by the presence of filaments, where $P_{\rm ram}$ can be up to two orders of magnitude larger than in the lowest density regions. Specifically, they find that filaments around low-mass structures have a Log($P_{\rm ram}$) ranging from $\sim 2 \times 10^{-13}$ to $\sim 6 \times 10^{-15}$~N~m$^2$ in in the region $1-3\times R_{200}$ from the centre of the group. 
It is therefore possible that some galaxies in groups and/or filaments experience gas stripping by ram-pressure. 

To assess whether our group galaxies are influenced by ram-pressure stripping, we must compare the expected ram-pressure with the self-gravity or anchoring pressure across the galaxy ($\Pi_{\rm gal}(r_{gal})$, where $r_{gal}$ is the radial distance from the galaxy centre), which reflects their ability to retain gas. Gas stripping will occur when $P_{\rm ram} > \Pi_{\rm gal}(r_{\rm gal})$.

We compute $\Pi_{\rm gal}$ for the group galaxies with a spiral morphology and a measured gas component (i.e. P5215 and P5055) using a pure disk model as described in \citetalias{Jaffe2018}, with the parameters listed in Table~\ref{tab:rp}, and assuming 
a gas fraction of 0.25 \citep[which corresponds to $\sim 10^{10}{\rm M_\odot}$ late-types,][]{Popping2014} and a disk scale-length for the gas $1.7$ times that of the stars \citep{Cayatte1994}. We also measured the extent of the \Ha emission disk along the semi major axis of the galaxies and computed $\Pi_{\rm gal}$ at that radius assuming this is the maximum radius at which ram-pressure has been able to strip gas. In other words, if we consider $r_{\rm extent}$ to be a truncation radius, $\Pi_{\rm gal}(r_{\rm gal}=r_{\rm extent})$ can be considered an upper limit to the $P_{\rm ram}$ experienced by the galaxies. Our results thus suggest that the maximum ram-pressure experienced by the two group galaxies is $P_{\rm ram}\leq 10^{-14}$N~m$^2$, which is consistent with the predicted $P_{\rm ram}$ from hydrodynamical simulations of groups and filaments above-mentioned, and even with the values of ram-pressure found in the outskirts of massive clusters. 

\begin{table*}
\caption{ Structural parameters and other physical properties of the gas-rich disk-dominated group galaxies: Disk to total light ratio (D/T), disk scale-length ($r_d$), disk mass ($M_{\ast_d}$),  extent of the H$_{\alpha}$ emission within the mayor axis of the disk ($r_{\rm extent}$),   projected phase-space coordinates,  and $\Pi_{\rm gal}$ at $r_{\rm extent}$.
\label{tab:rp}}
\centering
\begin{tabular}{lcccccccc}
\hline
ID    & D/T & $r_d$ &$\log (M_{\ast_d}/{\rm M_\odot}$) & $r_{\rm extent}$  	& r$/R_{200}$& v$/\sigma$ & $\Pi_{\rm gal}(r_{\rm gal}=r_{\rm extent})$	\\
      &     & (kpc) & &(kpc) 				& 	 		 & 			  & (N~m$^2$) 								   &   			\\
\hline
P5215 & $\sim$1	& 	2.3	& 10.18 & 12.6 & 	2.37	& 	-0.53	& $5.4\times10^{-15}$ \\
P5055 & $\sim$1	& 	4.2	& 10.25 & 15.0 & 	 	0.34	& 	-2.10	& 1.4$\times 10^{-14}$ 	\\
\hline\end{tabular}
\end{table*}


\subsection{Final remarks}


{  The galaxies presented in this work can be seen as representative 
of 
galaxies  at different stages in their journey from lower to higher density environments that will 
eventually quench.}

P5215 is representative of the galaxies that are about to fall onto relatively massive systems flowing through the filaments. These objects are still too far from the group center to strongly feel the relatively dense IGM of the groups, but still some effects of possible ram pressure stripping might be visible. These galaxies will most likely  get their halo gas removed by the sparse IGM of the filaments that is able to  compress the circumgalactic gas turning on the star formation in some clouds, which we observe with MUSE. We name  this process Cosmic Web Enhancement and to our knowledge no other study has reported this effect yet. In Paper XVI we present other pieces of evidence for this process. 

Galaxies falling onto groups with high relative velocities and on radial orbits get quite close to the group center, where the ram pressure stripping is effective. P5055 is one example. The truncated ionised disk that characterises the galaxy is a strong evidence in support to ram pressure stripping rather than strangulation. 
These galaxies will become quenched within few Gyr. 

Galaxies that instead fall onto groups on more tangential orbits never get to feel the pressure exerted by the hot IGM, but eventually quench for other reasons. P5169 is the GASP galaxy in the most advanced stage. It is now quenched and the analysis of the star formation histories show that the star formation declined throughout the disk in a similar way and on time scales of few Gyr. These pieces of evidence are consistent with strangulation. 

It is intriguing that we see evidence for strangulation in this galaxy, but see evidence for ram pressure stripping in P5055. There are three possibilities for why this would occur. Firstly, the galaxies could have different gravitational restoring forces, although they have similar stellar masses and morphology, suggesting this is not the case. Secondly, they could have different orbits, which would take them into different parts of the group with different relative velocities, leading to different ram pressure. This is very likely to be true given the wide range of orbits. However, there is a third possibility caused by the evolution of the group through time. P5055 appears to be infalling for the first time, but P5169 likely first encountered the group 2-4 Gyrs ago based on both its position within the backsplash region of group phase space and its decline in star formation history. Using the analytic mass accretion method of \cite{Correa2015}, we find that a group of mass 10$^{14.1}$ at z = 0 would have grown, on average, by 24$\%$ in the past 3 Gyrs. Given the X-ray scaling relations of \cite{Anderson2015}, this suggests that the density of the intragroup medium of this group would have increased by $\sim$ 30$\%$ over the same period. So, during its first interaction with the group, P5055 would have felt 30$\%$ more ram pressure than P5169 even if they had the same orbital parameters. Intriguingly, this may mean that we are witnessing the shift between strangulation and ram pressure stripping as a key quenching mechanism in this group.

Finally, the evolution of galaxies like P4946 is influenced by environmental effects acting on local scales, like gas accretion or mergers, rather than global effects. The fate of these fascinating objects is less predictable. 

\section{Conclusions}\label{sec:conc}
In this paper we have characterised the properties of a galaxy group in the local Universe and of its members. {  The group under investigation is unrelaxed and subject to contamination from filament members and projection effects; it also lacks of a characterisation at X-ray wavelengths, therefore its mass and density estimates are highly uncertain.} 

We have then focused on four group members that are part of GASP, an ESO Large program aimed at studying gas removal processes in nearby galaxies in different environments, with the intent of detecting signs of different physical processes at work on the ionised gas distribution  of galaxies in low mass systems. 

The main results can be summarized as follows:
\begin{itemize}
\item The group has not reached relaxation yet, and is found at the intersection of two filaments. It has a mass of 1.3 $\times 10^{14} {\rm M_\odot}$ and contains 28 members. Group members have rest frame (B-V) color and mass distributions skewed towards redder and more massive values than those of galaxies in other groups at similar redshift. The group contains galaxies of all morphologies and star forming properties. However, there is no a 1:1 correspondence between morphology, spectral type and color, indicating that the group has already undergone and/or is still undergoing a number of processes that affect galaxy properties on different time scales. 
\item Among the GASP galaxies, P5169 is the only passive one. It shows an early type morphology and  a regular stellar kinematics. Its position in the projected phase space diagram suggests the galaxy might have been in the group for a long time. However, P5169 completely quenched only a few $10^7$ yr ago. The shape of the star formation histories in different regions of the galaxy, the timescales at which quenching occurred and the symmetric fashion of the spatially resolved star formation histories suggest the galaxy has run out of gas due to strangulation that interrupted the fuel of the hot gas reservoir. This the first IFU observation of a post-strangulation galaxy.
\item P5215 is a face on spiral galaxy, 2.4 virial radii from the group center. It shows  regular stellar kinematics, while the ionised gas kinematics present some distortion in the outskirts. Specifically, it shows some detached clouds that suggest that either ram pressure stripping or cosmic web enhancement  due to its position within a filament are switching on star formation in dense circumgalactic clouds. One of these clouds is particularly peculiar:  it has $\sim 5$ kpc of diameter, mass of $10^8 {\rm M_\odot}$, total SFR of 0.06 ${\rm M_\odot} \, yr^{-1}$ located 27.5$^{\prime\prime}$ towards North. It  shows rotation, formed $\sim 2\times 10^7$ yr ago, has not stellar continuum detected. Given its distance, its systematically lower metallicity and its counter rotation with respect to P5215, the cloud is not a portion of the main galaxy that detached, nor is a low mass companion.
\item P5055 is a dusty edge on galaxy that present a ionised gas disk much smaller than the stellar disk. Gas and stars are characterised by similar kinematics. The disk started its depletion $\sim 6 \times 10^8$ yr ago. The galaxy is close to the core of the group traveling ant high speed, and all its characteristics are consistent with ram-pressure stripping.
\item P4946 is a barred S0 galaxy with an extended gaseous disk. This disk counter rotates with respect to the stellar disk, suggesting a different origin for the two components. Indeed it was formed only  $\sim 6 \times 10^8$ yr ago, while the stellar disk is much older (at least $\sim 10^{10}$ yr). Counter rotating disks are usually signs of recent past mergers or accretion. 
P4946 hosts an AGN in its core that, together with the presence of the bar, might be also responsible for the properties of the galaxy. 
\item Our analytical calculations support the scenario according to which ram pressure stripping might be the responsible for the features observed in the gas of P5125 and P5055. This is the first time the effect of the ram pressure is observed at optical wavelengths  in a group. 
\end{itemize}

These results show that spatially resolved data are the key to understand the origin and evolution of galaxies.


Our results emphasize the multitude of different physical processes taking place in {  low-density environments, such as forming groups and filaments, and their nd their influence on the star formation history of the galaxies. 
We present observational evidence of galaxies being
pre-processed in low-mass group and filament environment.
As the Universe evolves, these systems can eventually be accreted by a more massive halo, where galaxies will be further (and more violently) quenched. }

\section*{Acknowledgements}
We thank the referee, Prof. Harald Ebeling, for his insightful comments that allowed us to strengthen the results presented in this paper. 
Based on observations collected at the European Organisation for Astronomical Research in the Southern Hemisphere under ESO programme 196.B-0578. We acknowledge funding from the INAF PRIN-SKA 2017 program 1.05.01.88.04 (PI Hunt). 
Y.~J. acknowledges support from CONICYT PAI (Concurso Nacional de Inserci\'on en la Academia 2017) No. 79170132.
%



\bibliographystyle{mnras}
\bibliography{gasp}


\bsp	
\label{lastpage}

\end{document}